\documentclass[10pt,journal,compsoc]{IEEEtran}
\usepackage[disable]{todonotes}

\newcommand{\sketch}{\todo[color=yellow!30,inline]}    

\ifCLASSOPTIONcompsoc
  \usepackage[nocompress]{cite}
\else
  \usepackage{cite}
\fi
\ifCLASSINFOpdf
\else
\fi
\usepackage{svg}

\usepackage{cite}
\usepackage{amsmath,amssymb,amsfonts}
\usepackage{algorithmic}
 \usepackage{graphicx}
\usepackage{textcomp}
\usepackage{xcolor}
\usepackage{subcaption}
\usepackage[english]{babel}
\usepackage[utf8x]{inputenc}
\usepackage[T1]{fontenc}
\usepackage{subcaption}
\usepackage{mathtools}
\usepackage[linesnumbered,vlined,boxed,commentsnumbered, ruled]{algorithm2e} 

\usepackage{enumitem}
\usepackage{yfonts}
\usepackage{tikz}
\usetikzlibrary{shapes,snakes}
\newcommand*\circled[1]{\tikz[baseline=(char.base)]{
		\node[shape=circle,draw,inner sep=0.8pt] (char) {#1};}}
	
		\definecolor{customBlue}{RGB}{168, 218, 255}
	
	\newcommand*\circledBlue[1]{\tikz[baseline=(char.base)]{
			\node[shape=circle,draw,inner sep=0.8pt, fill=customBlue] (char) {#1};}}

\usepackage{pifont}
\usepackage{amsmath}
\DeclareMathOperator*{\argmin}{arg\,min}

\usepackage{svg}
\usepackage{graphicx}
\usepackage[colorlinks=true, allcolors=black]{hyperref}    

\usepackage{amssymb} 
\usepackage{tikz}
\usetikzlibrary{matrix}

\usepackage[absolute,showboxes]{textpos}
\TPMargin{5pt}
\textblockcolour{yellow!20}

\newcommand{\copyrightstatement}{
	\begin{textblock}{1}(0.00,0.0)    
		\noindent
		\scriptsize
		\copyright \  
		2020 IEEE.
Personal use of this material is permitted.  Permission from IEEE must be obtained for all other uses, in any current or future media, including reprinting/republishing this material for advertising or promotional purposes, creating new collective works, for resale or redistribution to servers or lists, or reuse of any copyrighted component of this work in other works.
		This paper is the accepted version to be published in IEEE Transactions on Dependable and Secure Computing. For the published version refer to DOI \url{https://doi.org/10.1109/TDSC.2020.3030605}
	\end{textblock}
}

\begin{document}
	\copyrightstatement
%
\title{AWARE: Adaptive Wide-Area Replication for Fast and Resilient Byzantine Consensus}
%
%
%
%

\author{Christian Berger,~
        Hans P. Reiser,~
        João Sousa,~
        and~Alysson~Bessani
\IEEEcompsocitemizethanks{\IEEEcompsocthanksitem C. Berger and H. P. Reiser are with Universität Passau, Germany.\protect\\
E-mail: \{cb,hr\}@sec.uni-passau.de
\IEEEcompsocthanksitem J. Sousa and A. Bessani are with LASIGE, Faculdade de Ciências, Universidade de Lisboa, Portugal.\protect\\
E-mail: \{anbessani,jcsousa\}@fc.ul.pt

}

\thanks{Manuscript accepted for publication Oct 04, 2020. Corresponding author: Christian Berger.
	For information on obtaining reprints of this article, please send an e-mail to:
	reprints@ieee.org, and reference the Digital Object Identifier below.
	Digital Object Identifier no. 10.1109/TDSC.2020.3030605}
}

%
%

\markboth{}{}
%



\IEEEtitleabstractindextext{%
\begin{abstract}
With upcoming  blockchain infrastructures, world-spanning Byzantine consensus is getting practical and necessary. In geographically distributed systems, the pace at which consensus is achieved is limited by the heterogeneous latencies of connections between replicas.
If deployed on a wide-area network, consensus-based systems benefit from weighted replication, an approach that utilizes extra replicas and assigns higher voting weights to well-connected replicas. This approach enables more choice in quorum formation and replicas can leverage proportionally smaller quorums to advance,  thus decreasing consensus latency.
However, the system
needs a solution to autonomously adjust to its environment if
network conditions
change or faults occur.
We present Adaptive Wide-Area REplication (AWARE), a mechanism that improves the  geographical scalability of consensus with nodes being widely spread across the world.
%
Essentially, AWARE is
 an automated and dynamic voting-weight tuning and leader positioning scheme, which supports the emergence of fast quorums in the system. It employs a reliable self-monitoring process and provides a prediction model seeking to minimize the system’s consensus latency.
In experiments using several AWS EC2 regions, AWARE dynamically optimizes consensus latency by self-reliantly finding a fast configuration yielding latency gains observed by clients located across the globe.

\end{abstract}

}

\maketitle
 \thispagestyle{plain}

\IEEEdisplaynontitleabstractindextext

%
\IEEEpeerreviewmaketitle

\IEEEraisesectionheading{\section{Introduction}\label{sec:introduction}}

%
%
%
%



\IEEEPARstart{S}{tate} machine replication (SMR) is a classical approach for building resilient distributed systems. It achieves fault-tolerance by coordinating client interactions with independent server replicas \cite{schneider1990implementing}.  Furthermore, SMR protocols typically use either some dynamically
selected leader~\cite{castro1999practical} or are fully decentralized~\cite{moniz2011ritas}.
In both cases, these  protocols usually require communication steps involving a
major subset of all nodes.

 With the emergence of novel Byzantine fault-tolerant (BFT) blockchain infrastructures, BFT SMR protocols have been increasingly catching academic attention over the last few years. For example, the BFT-SMaRt~\cite{bessani2014state} library has been employed as an ordering service~\cite{sousa2018byzantine} for the Hyperledger Fabric~\cite{androulaki2018hyperledger} blockchain platform allowing for a high-performance, resilient service execution that achieves sub-second latencies in a geographically distributed environment using the WHEAT~\cite{sousa2015separating} optimization. 
 \sketch{add more motivation for BFT in blockchains?}

In fact, BFT SMR protocols (typically called \emph{BFT consensus}) are a key component of permissioned blockchains as they can be used to establish total order of transactions in a closed group of processes without relying on the expensive proof-of-work mechanism~\cite{nakamoto2008bitcoin}, achieving thus a much higher performance even with few tens of nodes, a significant consortium size for this type of blockchain~\cite{sousa2018byzantine, androulaki2018hyperledger}.
However, they might still be deployed in a wide-area network (WAN) for geographic decentralization. 

\begin{figure}[t]
\centering
      \includegraphics[width=1\columnwidth]{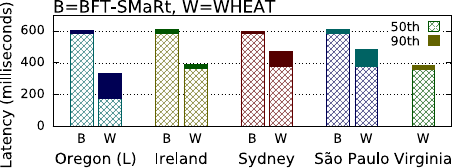}
\caption{Latency gain using weighted replication in
 WHEAT and BFT-SMaRt~\cite{sousa2015separating}, as measured by clients co-located with different replicas.}
\label{fig:wheat}
\end{figure}

\textit{Weighted replication.} While inter-node network latencies typically are similar between all pairs of nodes
in a LAN environment, we can observe variations of latencies in WAN environments.
The end-to-end latency of protocols waiting for a set of messages to be received
is determined by the slowest replica in a subset that forms a quorum, i.e., contains enough replicas to convince a replica to advance to the next protocol stage.
By introducing WHEAT~\cite{sousa2015separating} (WeigHt-Enabled Active replicaTion), Sousa and Bessani have shown that using additional spare
replicas and weighted replication allows the system to benefit from more variety in quorum formation. Thus, the system can make progress by accessing a proportionally smaller quorum of replicas and can obtain a significant latency decrease (see Figure~\ref{fig:wheat}). 

\begin{figure}[t]
	\centering
	\includegraphics[width=\columnwidth]{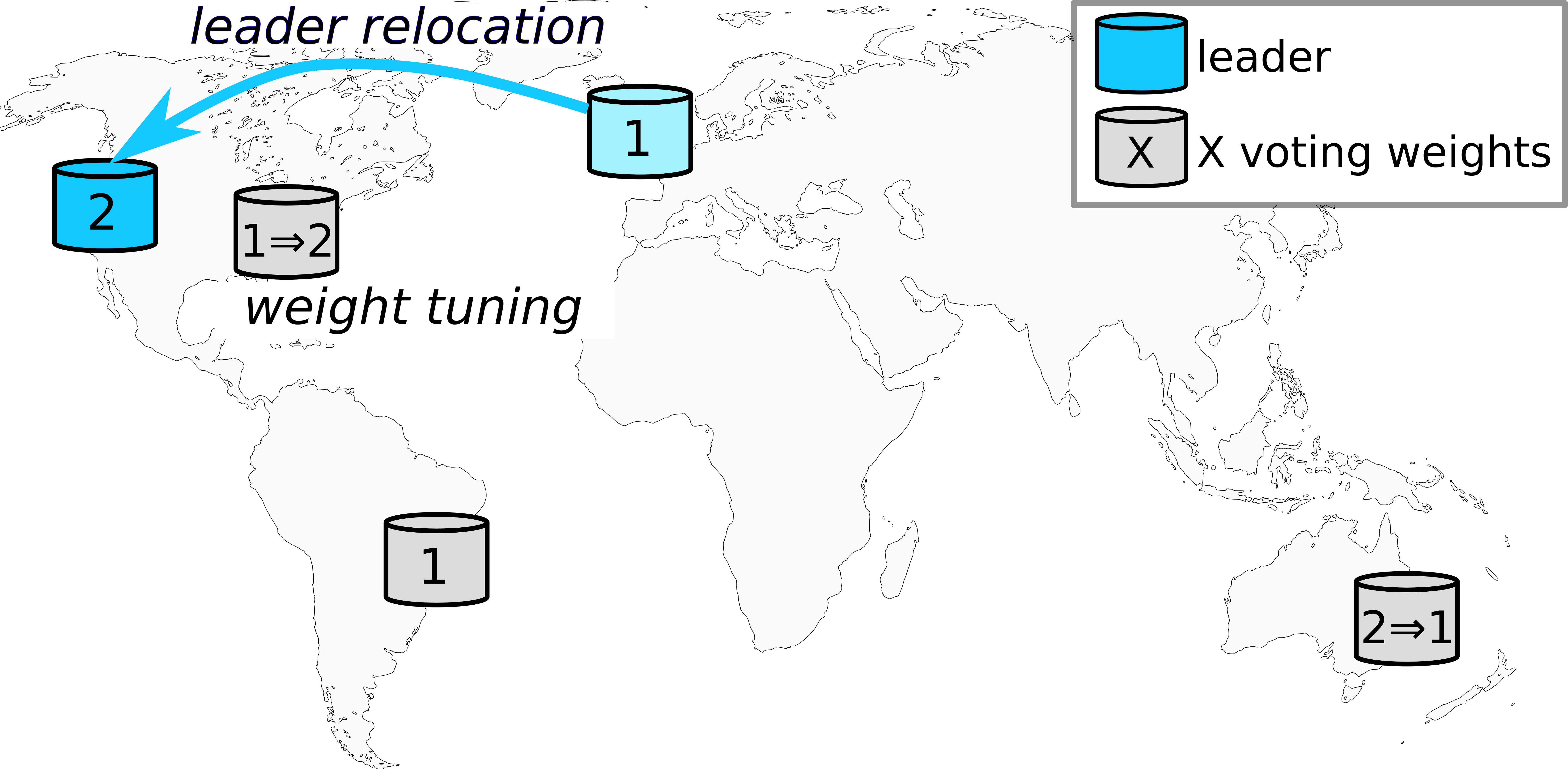}
	\caption{Optimizing a WHEAT configuration at run-time.}
	\label{fig:aware_map}
\end{figure}

\textit{Automation needed.} However, this benefit holds only if one selects the optimal weight configuration. A manual selection is difficult in practice, as the decision of what is the best
configuration for a given set of network characteristics is non-trivial.  Further,
network characteristics may be subject to run-time variations%
, and thus the optimal
configuration may also change at run-time.  To make weighted replication practically
usable, the SMR system needs a mechanism for automated and dynamic optimization. 
We present AWARE (Adaptive Wide-Area REplication), a mechanism for allowing geo-replicated state machines to self-optimize at run-time (see Figure~\ref{fig:aware_map}):  our method employs continuous self-measurements of the replicas' communication link latencies and analyzes the leader's expected consensus latency for different weight distributions and leaders.
It reconfigures the system to minimize consensus latency, 
thus also
 leading to latency gains observed by clients distributed across the globe.

\textit{Key constraints for SMR in blockchains.}
We want to tackle important challenges for SMR when used in blockchain systems, namely \textit{scalability} with respect to the number of nodes~\cite{vukolic2015quest}, \textit{geographical dispersion} as well as \textit{adaptivity} towards the environment. This paper extends our conference paper~\cite{berger2019resilient} by exploring how the scalability of the AWARE approach can be improved for larger systems. It also shows how AWARE can work as a resilient and adaptive ordering service for Hyperledger Fabric.

\subsection{Challenges and Contribution} We address practical challenges that arise when incorporating continuous self-optimization into WHEAT by proposing AWARE, 
which allows the SMR system to dynamically find a fast configuration and to adapt to changing environmental conditions during the system's life span. Our contributions aim for making WHEAT more viable for practical deployment.
Equipping a BFT system with such a self-optimizing mechanism introduces a set of challenges and research questions. In particular, our paper addresses the following problems:
\begin{itemize}
\item \textit{Self-monitoring in an SMR system.} Defining suitable strategies for self-monitoring, including the question of how the system can cope with incomplete and possibly counterfeit measurement information?

\item\textit{Deciding an optimal configuration.}
How can we compute the expected latency benefits for WHEAT configurations? We reason about consensus latency prediction and problems related to assessing configurations.
\item \textit{Safe and deterministic reconfiguration.}
In the overall algorithm, all correct replicas must reach agreement on a self-optimization before triggering a reconfiguration.

\item \textit{Mitigating faulty replicas' influence.} Can we redistribute high voting weights of unavailable (e.g., crashed) replicas and which threats impose malicious replicas?

\item \textit{Efficient self-adaption in larger systems.} How can we address a fast-growing configuration space (i.e., the problem of assigning weights) for larger systems?


\end{itemize}

%

\subsection{Outline}
\label{outline}
We start by explaining relevant preliminary work (§\ref{bft-replication}) and the system model we employ~(§\ref{system_model}). 
Then, we present AWARE's monitoring methodology (§\ref{measurements}) and subsequently describe our algorithm for finding an optimal configuration and deciding on a reconfiguration (§\ref{reconfiguration}). Moreover, we conduct an experimental evaluation of AWARE using different Amazon EC2 regions~(§\ref{evaluation}).
This paper is the extended version of \cite{berger2019resilient} and augments it by explaining how AWARE works in larger systems~(§\ref{larger_systems}) and how it is used as a self-adapting BFT ordering service for the Hyperledger blockchain platform~(§\ref{hyperledger}).
Finally, we discuss related research work (§\ref{related-work}) and draw conclusions (§\ref{conclusion}).

\section{BFT Replication}\label{bft-replication}
Castro and Liskov describe a practical replication algorithm for tolerating Byzantine faults, called Practical Byzantine Fault-Tolerance (PBFT)~\cite{castro1999practical},  that works in a weakly synchronous environment 
and incorporates optimization techniques to achieve throughput comparable to non-replicated services.  
Clients send requests to replicas (to the primary or upon timeout to all) that run consensus about the ordering of
requests. 
PBFT tolerates up to $f$ Byzantine servers, which may fail arbitrarily without compromising the service. The Byzantine fault model~\cite{lamport1982byzantine} usually requires a protocol to use a total of
$n = 3f + 1 $ replicas to guarantee both liveness and safety under the partially synchronous system model. 

\subsection{BFT-SMaRt}
BFT-SMaRt~\cite{bessani2014state} is an open-source library written in Java that implements robust and configurable BFT SMR. It has some important advantages compared to other SMR implementations -- for example,  UpRight~\cite{clement2009upright} or  PBFT~\cite{castro1999practical}: it employs a dynamically scalable replica set, 
provides a modular architecture using strictly separated and exchangeable components for different concerns, e.g., \textit{state transfer}, \textit{reconfiguration}, or \textit{consensus}, and it also achieves high performance because of its multi-core aware design and various optimization techniques it incorporates. Furthermore, it can be configured to run in crash fault tolerance (CFT) or BFT mode. For CFT, it requires fewer replicas ($n=2f+1$) and operates faster (2 protocol steps are then required for running consensus instead of 3).

The \textit{programming model} of BFT-SMaRt assumes the following request-response interaction model: Clients 
call the \textit{invokeOrdered(request)} interface to broadcast \textit{requests} 
 to all replicas.  The replicated server implements the deterministic \textit{executeOrdered(request)} interface, to which requests are delivered in \textit{total order},  thus ensuring that correct replicas process requests in the same sequence, and therefore also apply the same sequence of state transitions.
The client then waits for a specific quorum of matching replies. Replicas use a consensus algorithm to achieve the total order among all clients' requests: they decide the batch of requests to be executed next in every consensus instance. 
BFT-SMaRt uses the Mod-SMaRt protocol \cite{sousa2012byzantine}, an algorithm for SMR that employs an underlying leader-driven consensus primitive (the consensus algorithm described by Cachin \cite{cachin2009yet}). 

This Byzantine consensus algorithm consists of three phases~\cite{cachin2009yet}: \textit{PROPOSE}, \textit{WRITE}, and \textit{ACCEPT}. In the \textit{PROPOSE} step, the leader broadcasts a message that contains a batch of requests that need to be decided to all other replicas. The following two communication steps, \textit{WRITE} and \textit{ACCEPT}, are all-to-all broadcasts used for commitment. In these steps, each replica $i$ waits for a quorum $Q_i$ containing $\lceil\frac{n+f+1}{2}\rceil$ replicas to proceed (\textit{Byzantine majority}). Two different replicas $i \neq j$ might use two different quorums to advance, but these quorums overlap in at least one correct replica, i.e., $\vert Q_i \cap Q_ j  \vert \geq f+1 $. 



%

\subsection{WHEAT}
\label{WHEAT}

WHEAT\cite{sousa2015separating} is a variant of BFT-SMaRt's state machine replication protocol that is optimized for geo-replicated environments.
Its main innovation is the ability to decrease client latency by, counter-intuitively, \emph{adding} more replicas to the system.
The reason why this can result in a latency decrease, instead of the opposite, is because quorums in WHEAT are not formed using a Byzantine majority of replicas, as it is done in the rest of the BFT literature \cite{schneider1990implementing, castro1999practical, clement2009upright, sousa2012byzantine, EBAWA, amir10steward}.
In WHEAT's case, the size of a particular quorum can actually be smaller than a Byzantine majority.
Moreover, since WHEAT is expected to operate in wide-area networks, we can leverage the environment's heterogeneity to rely on the replicas that display the lowest end-to-end latency to be the ones forming these smaller quorums, and use the rest 
to form larger quorums that act as a fallback if the smaller ones are unavailable.

\begin{figure}[t]
\centering
\begin{subfigure}[t]{.22\textwidth}
   \includegraphics[width=1\columnwidth]{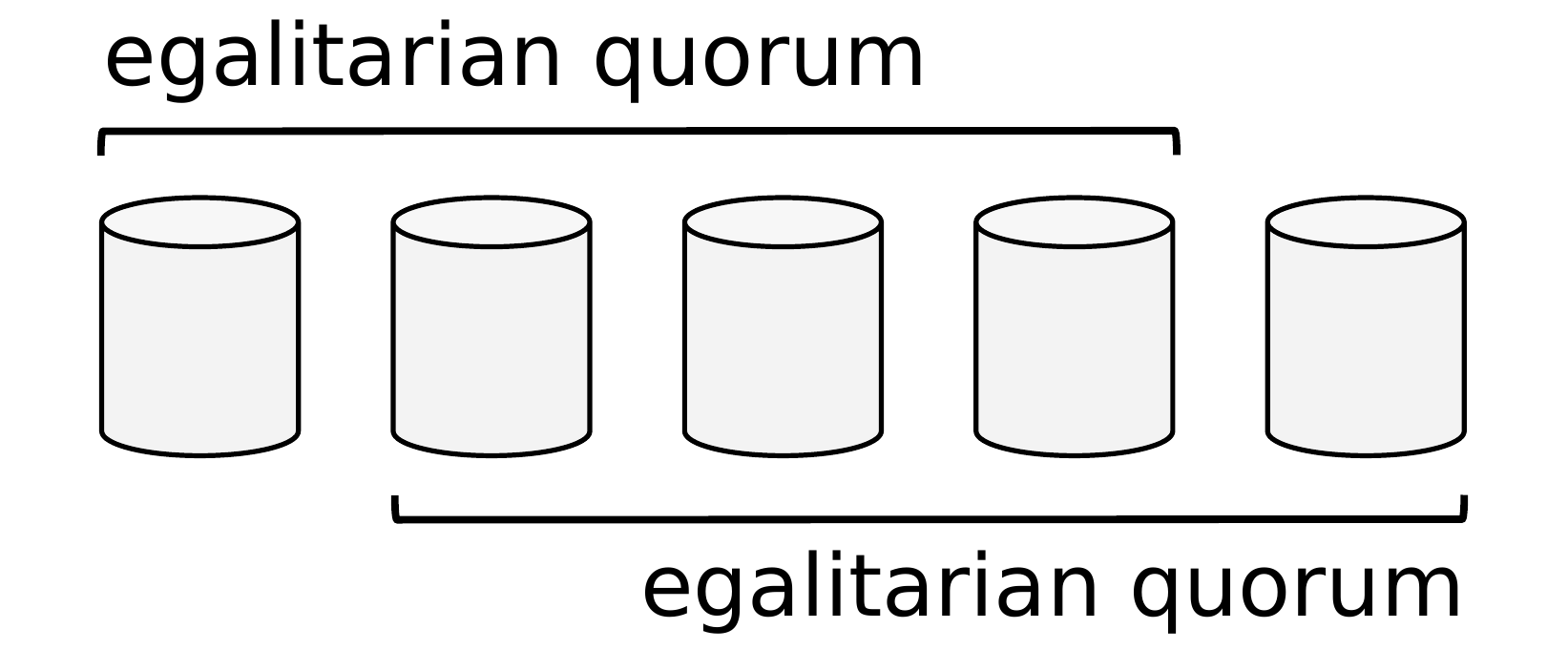}
   \caption{Egalitarian: all quorums contain $\lceil\frac{n+f+1}{2}\rceil$ replicas.}
   \label{fig:WHEAT_egalitarian}
\end{subfigure}
  \hfill
\begin{subfigure}[t]{.22\textwidth}
   \includegraphics[width=1\columnwidth]{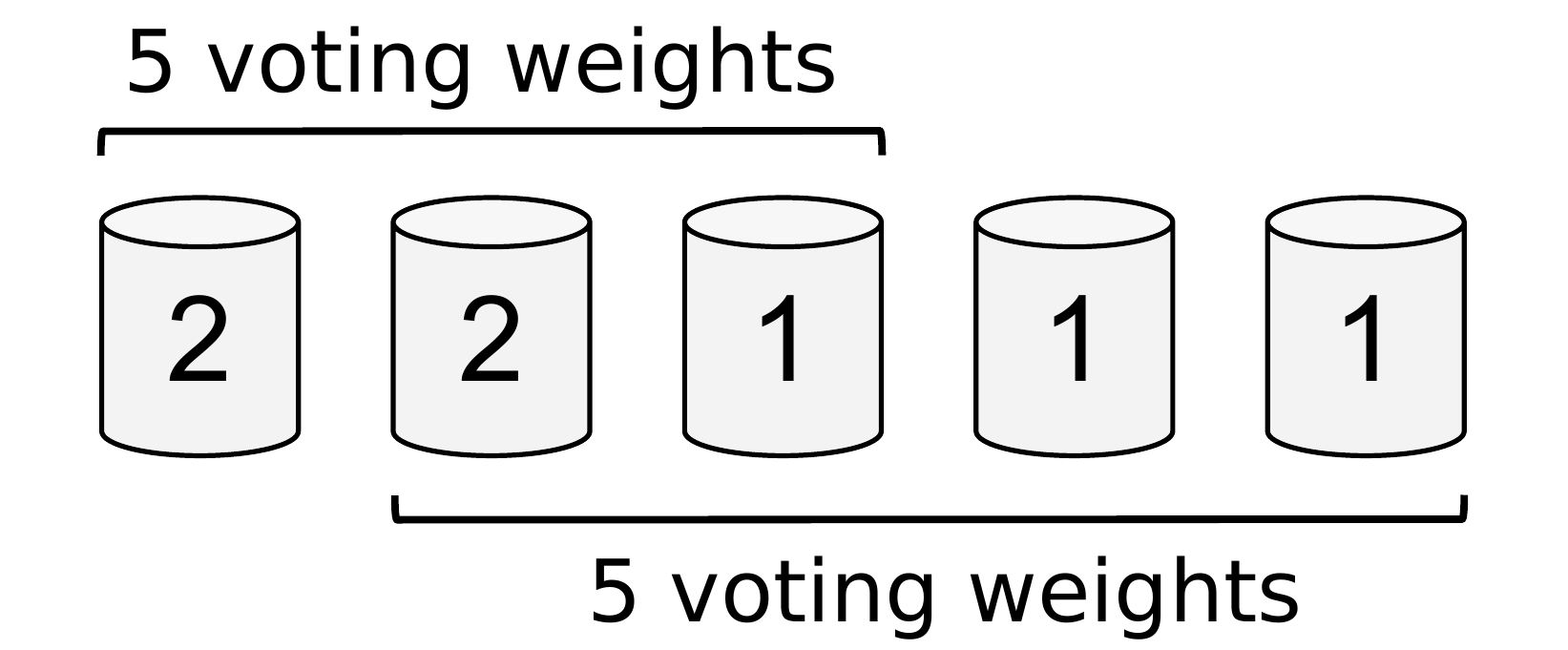}
   \caption{Weighted: a quorum contains at most $n-f$ and at least $2f+1$ replicas.}
\label{fig:WHEAT_weighted} 
\end{subfigure}
\caption[]{Possible quorums for $n=5$, $f=1$, $\Delta=1$ (BFT).}
\label{fig:WHEAT_quorums} 
\end{figure}

In order to understand how this works, let's consider a BFT system that, instead of  comprising the usual number of four replicas (the theoretical limit to withstand a single Byzantine fault),  actually comprises five (thus containing one extra replica).
Let's consider the quorum formation for this setup.
Recall that the definition of a BFT quorum system is a collection of subsets of replicas in which any two subsets intersect by $f+1$ replicas \cite{malkhi1998byzantine}.
To ensure quorum formation, BFT systems typically probe a Byzantine majority of replicas, as depicted in Figure \ref{fig:WHEAT_egalitarian}.
As we can see, using a Byzantine majority, the extra replica makes the quorum size increase from 3/4 to 4/5 across all possible combinations.
However, it is also possible to enforce quorum formation relying on weighted replication, as depicted in Figure~\ref{fig:WHEAT_weighted}.
In this case, by probing a majority of voting weights rather than a majority of replicas, we can see that there exist combinations of replicas that still intersect by $f+1$ replicas, thus forming quorums in the size of 3/5 and others in the size of 4/5.
Now imagine that this is a geo-replicated environment where the two best-connected replicas are assigned the highest weight with value 2.
In spite of having five replicas in the system, progress is made by typically probing three replicas to form a smaller quorum.
If for some reason any of these two fastest replicas is not available -- either due to a period of asynchrony or a failure -- progress can still be made by falling back to a quorum size of four replicas.
Moreover, we can also re-distribute weights if any of the preferred replicas become slower.
This approach is preferable to replacing replicas, since that would require new replicas to retrieve the state from others.

Further, for generalizing the above insight to any number of replicas, WHEAT employs the following safe weight distribution scheme~\cite{sousa2015separating}: let's assume a system of $n$ replicas, tolerating $f$ Byzantine faults and containing $\Delta$ additional replicas.
The relation between these variables is as follows:
    \begin{equation}
        n = 3f + 1 + \Delta
    \end{equation}
\normalsize
Moreover, to account for weighted replication, WHEAT demands each replica to wait for $Q_v$ voting weights, computed as follows:
    \begin{equation}
        Q_v = 2(f + \Delta) + 1
    \end{equation}
\normalsize
In order to correctly form quorums, WHEAT adopts a binary weight distribution in which a replica can have a value of either $V_{min}$ or $V_{max}$.
These values are computed as follows:
    \begin{equation}
        V_{min} = 1
    \end{equation}
		\begin{equation}
        V_{max}\footnote{$V_{max}$ can be a non-integer value.} = 1 + \frac{\Delta}{f}
    \end{equation}
\normalsize
Finally, $V_{max}$ is attributed to the $2f$ best-connected replicas in the system.
All other replicas are attributed $V_{min}$.
Using this distribution scheme, any quorum will contain between $2f + 1$ and $n-f$ replicas, instead of the fixed number of $\lceil\frac{n+f+1}{2}\rceil$ replicas as in traditional systems \cite{castro1999practical, bessani2014state}.


\subsection{Practical Implications}
\label{sect:resilience_performance_tradeoff}

\textit{Proportionally smaller quorums in WHEAT.} First, we would like to observe that the gains obtained by increasing $n$ and not $f$ in typical quorum-based protocols are not as good as in our approach. For example, the typical $f$-dissemination Byzantine quorum systems~\cite{malkhi1998byzantine} employed in most BFT protocols require quorums of $q=\lceil\frac{n+f+1}{2}\rceil$ replicas (for any $f$ and $n > 3f$), while in WHEAT/AWARE, the safe minimal quorum will always have $2f+1$ replicas, which is strictly smaller than $q$ for all $\Delta > 0$.

\textit{Resilience Performance Tradeoff.}
Second, the possibility to use additional replicas either to increase the $f$ bound of the system or $\Delta$, leads to a tradeoff situation between resilience and performance.
We consider $f$ a parameter of the system. Given a fixed value of $n$, e.g., the number of organizations running consensus nodes for a permissioned blockchain, we can trade off some resilience for smaller fast quorums, which may lead to better performance.


To illustrate this on an example, let's imagine an $n=13$ system: we could use the most resilient ($\Delta=0$, $f=4$) or a slightly less resilient but possibly better performing ($\Delta=3$, $f=3$) configuration. A configuration with higher $\Delta$ supports the emergence of fast quorums in the system. 

\begin{figure}[t]
	\centering
	\includegraphics[width=1\columnwidth]{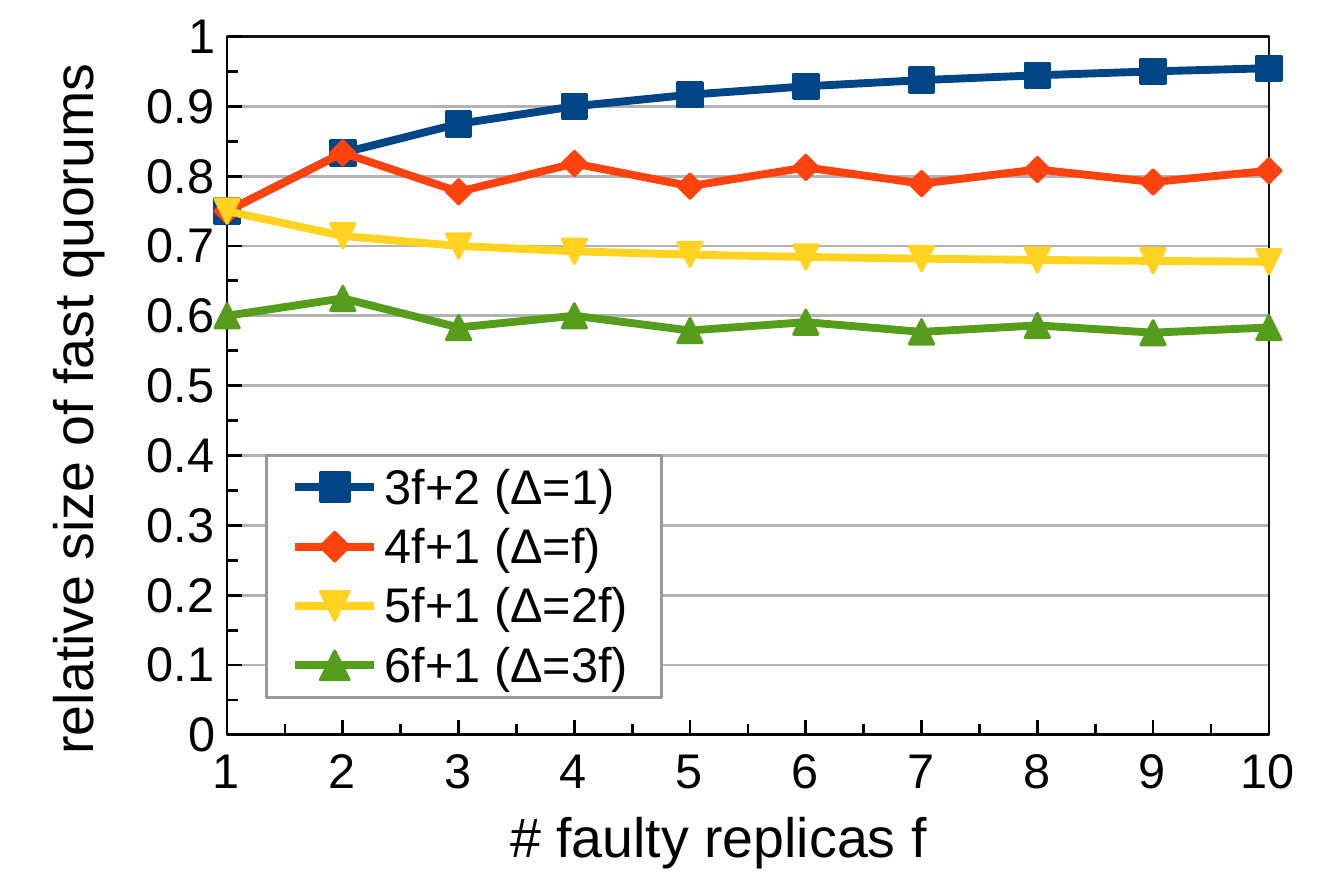}
	\caption{Relative size of fast quorums in proportion to the size of traditional Byzantine quorums for different system configurations.}
	\label{fig:size_fast_quorum}
\end{figure}

One may decide to choose the value of $\Delta$ depending on $f$. E.g., for $\Delta=f$, the proportion of faulty replicas that can be tolerated changes from roughly $\frac{1}{3}$ to $\frac{1}{4}$ of $n$. An interesting aspect of this is to have the values of $\Delta$ and $f$ be changed at run-time by an administrator, e.g., if changes in the threat level of the system are perceived.
Figure~\ref{fig:size_fast_quorum} illustrates the relative size of the small quorums consisting of the $2fV_{max}+1$ voting weights cast by $2f+1$ replicas to the traditional Byzantine quorums of size $\lceil\frac{n+f+1}{2}\rceil$ replicas for different system configurations.

\section{System Model}
\label{system_model}
We use the same system model as BFT-SMaRt and WHEAT, which we briefly summarize in this section.

\textit{Fault model.} The system consists of a replica set $I$ with a total of $n=3f+1+\Delta$ replicas and an arbitrary number of clients.
The number of faulty replicas is bounded by $f$, and we employ the Byzantine fault model for the behavior of faulty replicas. We guarantee that the safety of the system is never violated, while we require additional weak synchrony assumptions to guarantee liveness.



\textit{Communication model.} To guarantee liveness, we employ the partially synchronous system model~\cite{dwork1988consensus}, which is also popular with many other consensus-based systems, e.g., \textit{Steward}~\cite{amir10steward}, \textit{HotStuff}~\cite{hotstuff19} (a variant of which, \textit{LibraBFT}~\cite{baudet2019state}, is employed in Facebook's \textit{Libra} blockchain) and \textit{Stellar}~\cite{losa2019stellar}.
The system can be asynchronous initially, but we assume some upper bound $\delta$ on communication delay eventually exists after some unknown \textit{global stabilization time} ($GST$). Once $GST$ is reached, we can guarantee that the system makes progress. Note that we can always guarantee that the system stays safe -- even during periods of asynchrony. Communication is point-to-point, authenticated and reliable.
The implementation ensures this by using message authentication codes (MACs) over TCP/IP. The symmetric secrets for the replica-replica channels are established through \textit{Signed Diffie-Hellman} using an RSA key pair $(SK_i, PK_i)$ per replica.

\textit{Adversary model.}
The adversary has complete control over a set of replicas $B$ with $|B| \leq f$, where the adversary can freely choose replicas from the replica set $I$. For each $b \in B$, we consider the behavior of $b$ as \textit{arbitrary}, but still bound by computational capabilities, e.g., the attacker cannot have a malicious replica break strong cryptographic primitives (such as to circumvent authentication).
For instance, the attacker can let $b$ disregard incoming messages or inject messages into the network. 

In a worst-case scenario, an attacker chooses $f$ replicas with $V_{max}$ voting weights each, thus possessing in total $f V_{max}$ voting power. Even then, because of WHEAT's \textit{safe} voting scheme, there exists a fallback quorum of remaining $2f+1+\Delta$ replicas that have a majority of $fV_{max}+(f+1+\Delta)$ voting power. Under the assumption that replicas of the fallback quorum are correct, connected, and messages arrive within $\delta$, the system makes progress. Regardless of the last assumption, the system always remains safe.

We further assume that the attacker has the power to control and possibly disrupt the whole network but not for an indefinitely long time span and, as soon as sufficiently enough synchrony is reached (this point in time is modeled by $GST$), the system is guaranteed to make progress.



\section{Monitoring Strategy}
\label{measurements}

AWARE's self-optimization approach relies on sound self-monitoring capabilities of the system, which in turn require reliable measurements.
\subsection{Monitoring BFT Consensus}
\begin{figure}[tb]
\centering
   \includegraphics[width=1\columnwidth]{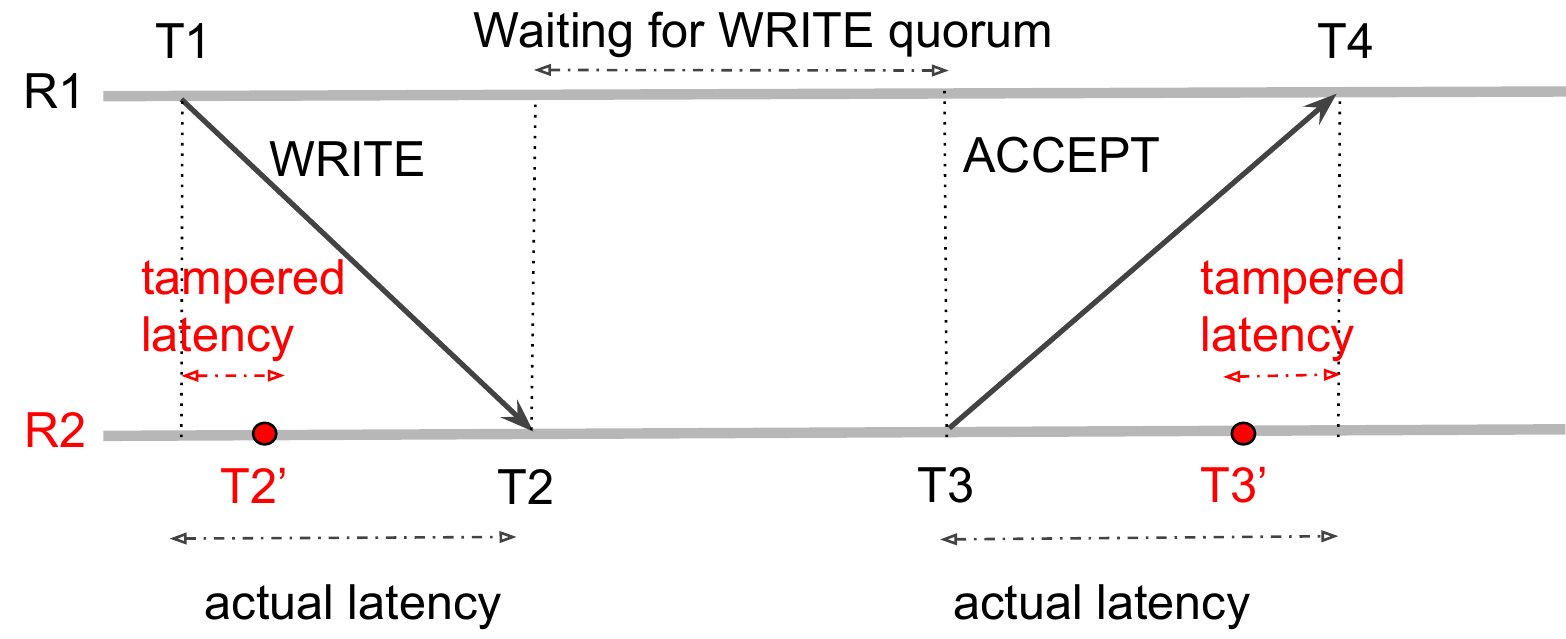}
   \caption{Problem with timestamps and Byzantine replicas.}
   \label{fig:measurements-BFT}
\end{figure}

\emph{Problem}.  A naive approach for monitoring the consensus pattern would be to observe
 the sending and reception time of WRITE and ACCEPT messages for all replicas.
Such \textit{quorum-based} measurements (e.g., measuring the time between replica $R_1$ sending a \textit{WRITE} to $R_2$ and receiving an \textit{ACCEPT} back from $R_2$) do not allow reasoning about link latencies because both message types \textit{do not casually depend on each other}. Replica $R_2$ might form a \textit{WRITE} quorum without $R_1$ and the \textit{WRITE} from $R_1$ might even arrive at $R_2$ after $R_2$'s \textit{ACCEPT} arrives at $R_1$.
 In a non-malicious setting, we could piggyback responses carrying timestamps $T_2$ and $T_3$  in \textit{WRITE} messages of subsequent protocol runs and thus compute link latencies using timestamps generated by both parties, e.g.,  by approximating the latency using $((T_4 - T_1) - (T_3 - T_2 ))/2$ as shown in Figure~\ref{fig:measurements-BFT}. However, malicious replicas can attach corrupt timestamps, e.g., malicious replica $R_2$ might try to shift $T_2$ closer to $T_1$ and $T_3$ closer to $T_4$, while correct replica $R_1$ has no means to detect this lying behavior and thus attributes $R_2$ a better latency. Byzantine replicas could try to abuse such behavior to increase their voting weights.

%
%
%

\emph{\textit{WRITE} and \textit{ACCEPT}.} 
To prevent this, we favor one-sided measurements, which require only \textit{the measuring replica to be correct}. In this method, we employ additional response messages for monitoring the consensus pattern: replicas \textit{immediately} respond to a protocol message by directly sending a \textit{WRITE}-\textit{RESPONSE} after receiving a \textit{WRITE}. Let $T'_4$ denote the arrival of the response message back at the measuring replica.
Then, the measuring replica can use $(T'_4-T_1)/2$ as  one-way link latency. 
However, this introduces monitoring overhead into the system. 
These latencies allow us to reason about times in which replicas form weighted quorums to proceed to subsequent protocol stages.

\emph{\textit{PROPOSE}.} 
Proposals are larger than other messages because they carry the actual consensus proposal (a batch of client requests) instead of just a cryptographic hash, thus they may have higher latency. 
This is also relevant for predicting the consensus latency, because every replica can only start broadcasting its \textit{WRITE} message after having received a \textit{PROPOSE} first. Further, we also use a response message for the latency measurement of this phase. 


The \textit{PROPOSE} latency is also relevant for \textit{automated leader location optimization}. Therefore, we measure the latencies of non-leaders proposing to other replicas 
 in order to determine hypothetical latency gains for the system when using a different replica as the protocol leader.  



\subsection{AWARE Approach}
\label{variants}
Following a systematic approach, we develop customizable variants of AWARE. 
 In the following, we give a brief summary of design decisions and configurable options.
 
\emph{Response to \textit{WRITE}.} In our approach, we expect each correct replica $i$ to measure the latencies of its point-to-point links to every other replica and maintain a latency vector $L_i =  \langle l_{i,0}, ..., l_{i,n-1} \rangle $. 
We use the \textit{WRITE}-\textit{RESPONSE}\footnote{We do not need to use an additional \textit{ACCEPT-RESPONSE} because the \textit{ACCEPT} phase has the same message pattern as the \textit{WRITE} phase.} messages to measure latencies between replicas. Further, the response message needs to include a challenge, e.g., a number that was beforehand randomly generated by the sender and attached to the original protocol message. This way  we can guarantee that a replica has received the \textit{WRITE} and that Byzantine replicas cannot send responses to messages before actually having received them.


\emph{Non-Leaders' \textit{PROPOSE}.} The \textit{DUMMY}-\textit{PROPOSE} allows measuring precisely the time non-leaders need to \textit{PROPOSE} batches of possibly large size to the rest of the system, where we expect a difference in cases where the network becomes the bottleneck. 
 Non-leaders do not simultaneously propose to avoid creating a high overhead to the system,
 degrading its performance and counter-acting our goal of improving the performance. 
 We use a \textit{rotation scheme} in which only one additional replica simultaneously broadcasts a \textit{DUMMY}-\textit{PROPOSE} along with the leader, proposing a dummy batch in the same way as the leader does, but without starting a new consensus instance, and all replicas  disregard the proposal. Replicas reply with a \textit{PROPOSE}-\textit{RESPONSE} 
and include the proposed batch in the response message to ensure symmetric behavior on communication link latency.
Using the \textit{DUMMY}-\textit{PROPOSE} is optional as it introduces overhead to the system (see §\ref{throughput}) and it is also possible to approximate these latencies using the measurements of \textit{WRITE}-\textit{RESPONSE}. 
\begin{figure}[tb]
\centering
      \includegraphics[width=0.9\columnwidth]{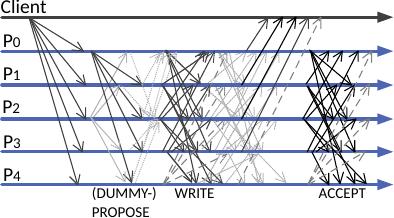}
\caption[]{Message flow of AWARE ($f=1;\Delta=1$). 
}
\label{fig:message-flow-AWARE} 
\end{figure}

Figure \ref{fig:message-flow-AWARE} shows the \emph{message flow}\footnote{The message pattern of WHEAT/AWARE differs from BFT-SMaRt in the use of \textit{tentative executions}, an optimization that was introduced in PBFT~\cite{castro1999practical}. } of AWARE  utilizing all monitoring messages. This yields the variant of AWARE with the highest accuracy in leader selection.
Furthermore, AWARE defines the number of recent monitoring messages to be used for computation of the  latencies for each connected replica in a configurable \emph{monitoring window}.

Moreover, in AWARE each correct replica $i$ periodically reports its latency vector $L_i$ to all other replicas. Replicas do this after some configurable \emph{synchronization period} by disseminating these measurements with total order (ordered together with the client requests) so
 that all replicas maintain the same latency matrix after some specific consensus instance. 
We employ a deterministic procedure for deciding a reconfiguration and use the same monitoring data in all correct replicas (while it would also be possible for replicas to have distinct views on the measurements and then run consensus on possible actions).

Once replicas have synchronized measurements after a given consensus instance, they employ the model we explain in §\ref{latency-prediction} to predict the best weight distribution and leader. Replicas use a \emph{calculation interval} $c$ defining the number of consensus instances after which a calculation and possibly a reconfiguration is triggered.

\emph{Bounding monitoring overhead}. We can arbitrarily decrease the monitoring overhead by specifying a  parameter $\omega~\in~[0,1]$ that determines the  maximum overhead induced by the monitoring procedure. Implementation-level details such as the frequency of sending specific monitoring messages (e.g., \textit{DUMMY}-\textit{PROPOSE}) are automatically derived from $\omega$.
Frequent measurements provide more up-to-date monitoring data and allow for faster reaction to environmental changes but also negatively impact the maximum throughput (see §\ref{throughput}).

\subsection{Sanitization}
\label{sanitization}
All replicas maintain synchronized latency matrices $M^{P}$ and $M^{W}$ for keeping measurements of \textit{PROPOSE} and \textit{WRITE} latencies,  both initially filled with entries
    \begin{equation}
M^{x}[i,j]\leftarrow  \begin{cases}
    +\infty ,& \text{if } i \neq j\\
   0 ,             & \text{otherwise}
\end{cases}
    \end{equation}
\normalsize
for $x \in \{P,W\}$. $M^x[i,j]$ expresses the latency of replica $i$ to $j$ as measured by $i$ for message type $x$ (either \textit{PROPOSE} or \textit{WRITE}). Further, replica $i$ can update a row of these matrices with its measurements $L_i^{P}$ and $L_i^{W}$ 
by using the \textit{invoke} interface of BFT-SMaRt: 
$$
\textit{invokeOrdered($\mathsf{MEASURE},L_i^{P},L_i^{W}$)};
$$
\normalsize
With this interface of BFT-SMaRt, we create a global total order on all client requests \textit{and} measurement messages.
The updating process yields a matrix $M^x$, with $M^x[i,j] = L^x_i[j]$ if replica $i$ sent its measurements within the last calculation interval $c$ of measurement rounds, or a missing value ($+\infty$) if it did not send any measurements within the last $c$ rounds.

\begin{figure}[tb]
	\definecolor{yellowgreen}{RGB}{191, 255, 0}
	\centering
	\begin{subfigure}[b]{.235\textwidth}
		\centering
		\footnotesize 
		\[
		\tikz [baseline=(m.center)]
		\matrix (m) [
		matrix of math nodes,
		left delimiter={[},
		right delimiter={]},
		column sep=-0.1pt,
		row sep=-0.1pt,
		nodes={minimum width=2.4em, inner sep=2pt},
		r/.style={fill=red!20},
		o/.style={fill=yellowgreen!30},
		g/.style={fill=green!30}
		] {
			0               & |[g]| 68               & |[g]| 69              & |[g]| 93                 & |[g]| 40                \\
			|[o]| 67              & 0                & 	|[g]| 133             & |[g]| 92                 & |[g]| 35                \\ 
			|[g]| 69              & 	|[o]| 132              &  0               & |[g]| 157                & |[g]| 99              \\ 
			|[o]| 92              &  |[g]| 92              & |[o]| 156             &  0                  & |[g]| 70                \\ 
			|[r]| 0          & |[r]| 0              & |[r]| 0             & |[r]| 0                &  0    \\
		};
		\]
		\caption{Before sanitization.}
	\end{subfigure}
	\begin{subfigure}[b]{.235\textwidth}
		\centering
		\footnotesize 
		\[
		\tikz [baseline=(m.center)]
		\matrix (m) [
		matrix of math nodes,
		left delimiter={[},
		right delimiter={]},
		column sep=-0.1pt,
		row sep=-0.1pt,
		nodes={minimum width=2.4em, inner sep=2pt},
		r/.style={fill=red!20},
		o/.style={fill=yellowgreen!30},
		g/.style={fill=green!30}
		] {
			0               & |[g]|68               & |[g]|69              & |[g]|93                 & |[g]|40                \\
			|[g]|68              & 0                & |[g]|133             & |[g]|92                 & |[g]|35                \\ 
			|[g]|69              & |[g]|133              & 0               & |[g]|157                & |[g]|99              \\ 
			|[g]|93              & |[g]|92              & |[g]|157             & 0                  & |[g]|70                \\ 
			|[g]|40              & |[g]|35               & |[g]|99             & |[g]| 70                 &  0   \\             
		};
		\]
		\caption{After sanitization.}
	\end{subfigure}
	\caption{Sanitized matrix of median \textit{WRITE} latencies.  
	}
	\label{fig:mattrices}
\end{figure}

We sanitize both matrices immediately before the calculations happen to mitigate the influence of malicious replicas.
We do that by exploiting the symmetry characteristic of replica-to-replica latencies and let replicas have a pessimistic standpoint on measurements.  They use the pairwise larger delay in calculations so that replicas cannot make themselves appear faster. This procedure yields 
    \begin{equation}
\hat{M}^x[i,j] = \mathit{max}(M^x[i,j],M^x[j,i])
\end{equation}
\normalsize
for $x \in \{P,W\}$. $\hat{M}^P$ and $\hat{M}^W$ are then the sanitized latency matrices for \textit{PROPOSE} and \textit{WRITE}, respectively.  Figure~\ref{fig:mattrices} presents an example for sanitization.
This way, Byzantine replicas cannot fraudulently improve their link latency to any \textit{correct} replica, and they also cannot blame (worsen a link latency to) a correct replica \textit{without being contributed a bad latency themselves}. Still, in case of $f>1$, Byzantine replicas may show intriguing behavior, e.g., by claiming to have tremendous connections to each other (see §\ref{analysis}).

%

\section{Self-optimization}
\label{reconfiguration}
In our self-optimizing approach, replicas deterministically reconfigure to a new weight configuration and/or leader position. This requires all replicas to (1) agree on what the optimal configuration is and (2) decide whether they will adjust themselves accordingly by triggering a \textit{view change}. 
\subsection{Optimizations}
Overall, AWARE employs two  dynamic optimizations:

\emph{Voting weights tuning.}
Adjusting voting weights leads to latency gains observed by clients across all sites~\cite{sousa2015separating}. AWARE searches for a weight distribution that optimizes the system's consensus latency.

\emph{Leader relocation. }
Relocating the leader in a well-connected site of the system reduces the request latency observed by clients~\cite{sousa2015separating, liu2017leader}. Hence, AWARE is capable of selecting the leader location as an optional optimization technique. 

We follow the idea of obeying \emph{leader selection constraints}. AWARE employs an abstraction where the BFT protocol provides an interface for choosing a leader. In particular, we allow the protocol to provide a suitable set of \textit{leader candidates} $\mathfrak{L}$ out of which AWARE chooses one.

In context of their decision, an \emph{optimization goal} $\alpha$ defines the threshold (relative to the current configuration) by which a predicted consensus latency needs to be faster to trigger a reconfiguration. This prevents the system from \textit{jumping back and forth} between configurations that are almost equally fast.

Further, AWARE optimizations operate at the granularity of logical protocol rounds that successfully lead to an agreement, called \textit{consensus instances}.
Reconfigurations are evaluated at most once every \textit{calculation interval}\footnote{The calculation interval determines the frequency of optimizations and can be configured in our implementation. Our default value is 500.} $c$ of consensus instances is reached. This is to ensure the
stability of our approach by not performing reconfigurations too frequently. 


\subsection{Consolidated Protocol}
We ensure determinism by providing a deterministic consensus latency prediction function used for our calculation (see Algorithm~\ref{algo:PredictLatency}). Further, we ensure that the measurement data is the same in all replicas after a specific consensus instance by synchronizing measured latency information with total order broadcast. Replicas apply a change to the \textit{current view} object to update the weight distribution and leader after a calculation that is performed in intervals
specified by logical protocol rounds (consensus instances) rather than by time.
This works as follows:

\sketch{Maybe TODO: Rewrite the following, into subsubsections:  Consolidating Measurements (1-5), Finding an optimal configuration (6),  Reconfiguration (7))}
\begin{enumerate}[leftmargin=14pt, label=\arabic*.]

\item Each replica $i$ collects its latency measurements (a moving median) in a vector $L_i = \langle l_{i,0}, ..., l_{i,n-1} \rangle$;

\item Periodically, each replica $i$ disseminates its vectors for \textit{PROPOSE} ($L_i^{P}$) and \textit{WRITE} ($L_i^{W}$) with total order by calling $\mathit{invokeOrdered(\mathsf{MEASURE},L_i^{P},L_i^{W})};$

\item Once a replica $i$ decides a batch, that batch may contain messages $\langle \mathsf{MEASURE},L_j^{P},L_j^{W} \rangle$ from some replica $j \in I$. It uses these vectors to update its synchronized matrices $M^P_i$ and $M^W_i$,  i.e., for replicas $k=0,..,n-1$ assigning $M^W_i[j,k]=L_j^{W}[k]$ that is the information that $i$ has about the latency between replica $j$ and all other replicas measured by $j$.
This applies to the maintained latency information for both \textit{PROPOSE} ($M^P_i$) and \textit{WRITE} ($M^W_i$). 

\item When a defined number (specified by the calculation interval $c$) of consensus instances is reached, all replicas have the same matrices $M^P$ and $M^W$, e.g., with $M^W[i,j] =L^W_i[j]$ if replica $i$ sent its \textit{WRITE} measurements within the last $c$ consensus instances, or $M^W[i,j] = +\infty$ if $i$ did not send its measurements. The same applies to  $M^P$.

\item The next step is to deterministically sanitize the matrices to avoid the influence of malicious replicas (see §\ref{sanitization}), generating $\hat{M}^P$ and $\hat{M}^W$.

\item Now, every replica solves the following optimization problem, where $\mathit{PredictLatency}$ (Algorithm~\ref{algo:PredictLatency}) is a function for predicting the latency of the consensus protocol  using the latencies in $\hat{M}	^P,\hat{M}^W$, and a set of weight distributions $W \in \mathfrak{W}$ and permitted leaders $l \in\mathfrak{L}$: 
\begin{equation}
 \langle \hat{l}, \hat{W} \rangle =\argmin\limits_{W \in \mathfrak{W}, l \in \mathfrak{L}} \mathit{PredictLatency}(l,W,\hat{M}^P,\hat{M}^W) 
\end{equation}
In the end, the configuration $\langle \hat{l}, \hat{W} \rangle$ that provides optimal leader consensus latency is the one selected for the next reconfiguration if the predicted latency is better than the current configuration by the factor $\alpha$ (optimization goal). Note that since this procedure is deterministic, the $\langle \hat{l}, \hat{W} \rangle$ is the same in all replicas.

\item In case the replicas find a faster weight configuration, they  update their view to respect the new voting weights for the following consensus instances. Optionally, if the system uses leader relocation, the replicas might also trigger a view change to elect a faster leader.
\end{enumerate}

\subsection{Latency Prediction}
\label{latency-prediction}
We predict the optimal configuration by simulating a protocol run for each configuration using the sanitized latency matrices
$\hat{M}^{P}$ and $\hat{M}^{W}$
to compute the predicted consensus latency of the leader replica and subsequently select a configuration that minimizes this latency.  We argue that a fast configuration is one that \textit{yields a low consensus latency from the perspective of the leader.} Once having finished a consensus instance, a leader can prepare and propose the next batch~\cite{sousa2012byzantine, bessani2014state}, while, in the meantime, poorer connected replicas may still wait for messages to form their quorums. To make progress, a leader needs to fulfill its quorums of $Q_v$ voting weights as well, so it needs to be well connected to replicas with preferably high voting weights.  The leader itself is always assigned
$V_{max}$ voting weights in order to minimize its consensus latency.

\begin{algorithm}[hbt] 
	\footnotesize
	\caption{\emph{formQV} computes the times  replicas form weighted quorums of $Q_v=2fV_{max}+1$ voting weights. 
	}
	\DontPrintSemicolon
	\KwData{replica set $I$, latency matrix $\hat{M}^{W}$,   times $(T_{i}^{\mathit{current}})_{i \in I}$ and voting weights $(V_{i})_{i \in I}$}
	\KwResult{times $(T_{i}^{\mathit{next}})_{i \in I}$ replicas can advance to the next protocol stage}
	\label{algo:formWeightedQuorum}
	
	\For{ $i \in  I$}{
		$\mathit{received_i} \leftarrow  \text{new PriorityQueue()}$ \;
		\For{ $j \in  I$}{
			$\mathit{received_i}.add(\langle T_j^{\mathit{current}}+\hat{M}^{W}[j,i], V_j  \rangle)$\;
		}
	}
	\For{ $i \in  I$}{
		$  \mathit{weight} \leftarrow  0$ \;
		\While{$\mathit{weight} < Q_v$}{
			$\langle T_{\mathit{next}}, V_{\mathit{next}}  \rangle \leftarrow \mathit{received_{i}}.\mathit{dequeue()} $ \;
			$\mathit{weight} \leftarrow \mathit{weight} + V_{\mathit{next}}$  \;
			$ T_i^{\mathit{next}} \leftarrow T_{\mathit{next}} $ \;
		}
	}
	
	\textbf{return} $(T_{i}^{\mathit{next}})_{i \in I}$ \;
\end{algorithm}

\begin{algorithm}[hbt] 
	\footnotesize
	\caption{\emph{PredictLatency} computes the consensus latency (amortized over multiple rounds)}
	\DontPrintSemicolon
	\KwData{replica set $I$, leader $p$, system sizes $n$, $f$, $\Delta$, weight config.  $W = \langle R_\mathit{max}, R_\mathit{min}
		\rangle$, latency matrices for \textit{PROPOSE} $\hat{M}^{P}$ and \textit{WRITE}  $\hat{M}^{W}$,  consensus rounds $r$ }
	\KwResult{consensus latency of the AWARE leader}
	\label{algo:PredictLatency}
	$V_{\mathit{max}} \leftarrow 1 + \frac{\Delta}{f}$ \ \
	$V_{\mathit{min}} \leftarrow 1$  \ \
	$V_i \leftarrow 
	\begin{cases}
	V_{\mathit{max}},& \text{if } i \in R_{\mathit{max}}\\
	V_{\mathit{min}},              & \text{otherwise}
	\end{cases} $ \;
	$\forall i \in I: \ \mathit{offset}_i \leftarrow 0 $   \\ 
	\While{$r > 0$} {    
		\For{ $i \in  I$}{ 
			$T_i^{\mathit{PROPOSED}} \leftarrow  max(\hat{M}^{P}[p,i], \mathit{offset}_i ) $ \;
		} 
		$({T_i^{\mathit{WRITTEN}}})_{i\in I} \leftarrow formQV(I, \hat{M}^{W}, (T_{i}^{\mathit{PROPOSED}})_{i \in I}, (V_{i})_{i \in I})$ \;
		$ (T_{i}^{\mathit{ACCEPTED}})_{i \in I} \leftarrow formQV(I, \hat{M}^{W}, (T_{i}^{\mathit{WRITTEN}})_{i \in I}, (V_{i})_{i \in I})$ \;
		\For{$i \in I$}{
			$\mathit{offset}_i \leftarrow T_i^{\mathit{ACCEPTED}} - T_p^{\mathit{ACCEPTED}} $ \;
		}	
		$\mathit{consensusLatencies}_{r}  \leftarrow T_p^{\mathit{ACCEPTED}}$  \;
		$r \leftarrow r - 1$ \;
	}
	\textbf{return} \text{average of} $\mathit{consensusLatencies}$ \;
\end{algorithm}

\begin{figure}[tb]
	\centering
	\includegraphics[width=1\columnwidth]{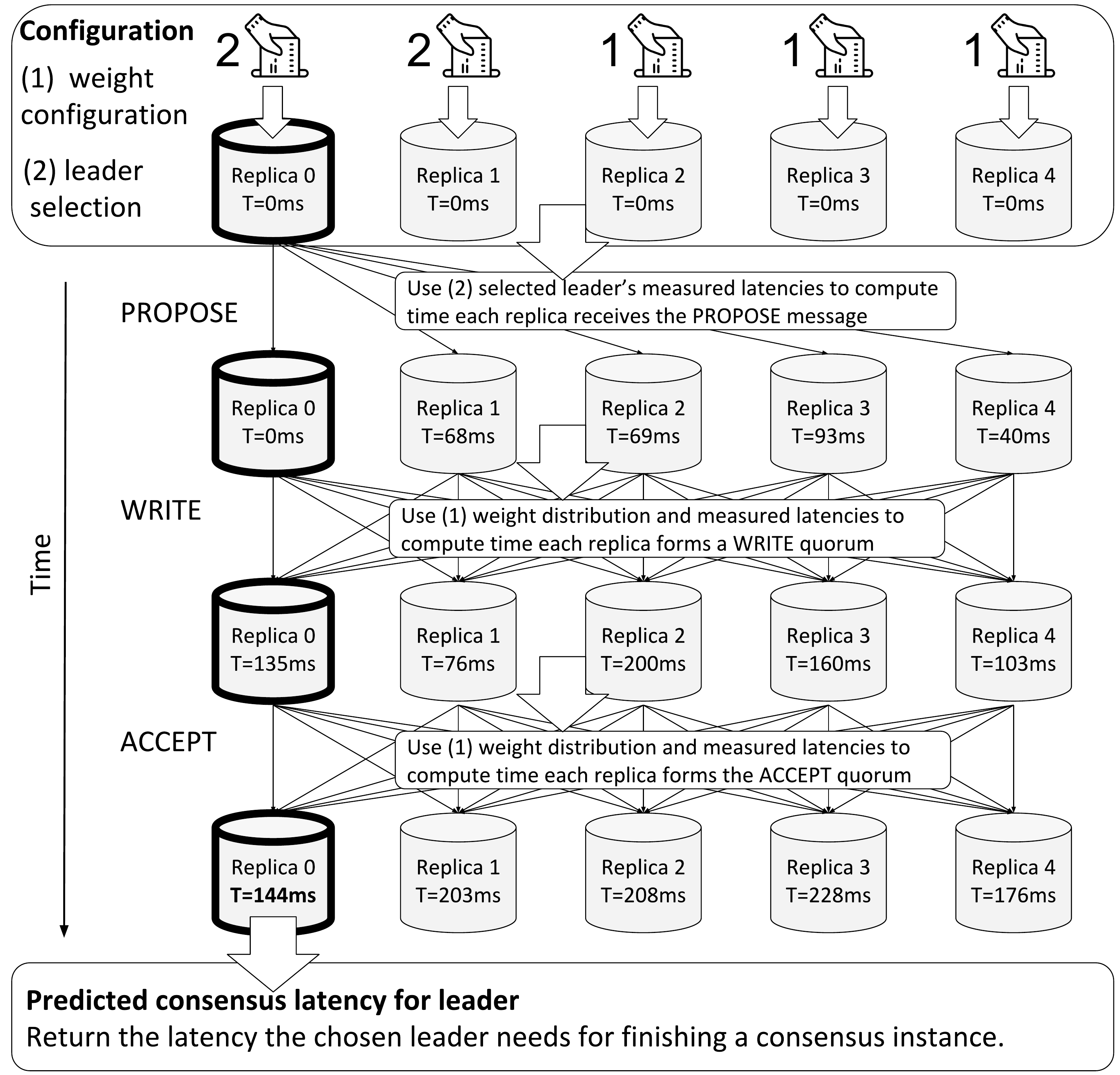}
	\caption{Computing the latency of a WHEAT consensus (here:  $f=1,\Delta=1$) for a given  configuration.}
	\label{fig:LatencyPrediction}
\end{figure}

Algorithm~\ref{algo:formWeightedQuorum} computes, for each replica, the time it can proceed to a subsequent protocol stage by forming a weighted quorum of $Q_v = 2V_{max} + 1$ voting weights given the time a replica starts in its current stage $T^{current}_{i \in I}$ and latency matrix $\hat{M}^{W}$. In particular, the algorithm computes the time a replica receives the \textit{WRITE} messages from all others (lines 1--4) and then computes for every replica the time it has gathered enough voting weights $Q_v$ to, e.g., proceed from \textit{WRITE} to \textit{ACCEPT} phase (lines 5--10). This works by letting each replica pull from a priority queue, in which incoming messages are sorted by ascending arrival time, and accumulate the voting weights until $Q_v$ is reached.  The arrival time of the last message, necessary to reach  this quorum, determines the time a replica can proceed to the next protocol stage.

Algorithm~\ref{algo:PredictLatency} is used to predict the leader's consensus latency for a given configuration by simulating the consensus protocol run. It first computes the times each replica receives the leader's proposal (line 4--5), then
 it uses Algorithm~\ref{algo:formWeightedQuorum} as building block to compute the times replicas complete the \textit{WRITE} (line 6) and \textit{ACCEPT} (line 7) stage, respectively.  
Further, this algorithm computes the amortized leader consensus latency over multiple rounds $r$. 
Note that replicas achieve consensus at different times (as can be seen in Figure~\ref{fig:LatencyPrediction}) and if the difference between the time leader $p$ decides and the time replica $i$ decides is greater than the propose latency $\hat{M}^{P}[p,i]$, then replica $i$ might receive a \textit{PROPOSE} for the next consensus instance but wait for its last consensus to finish before broadcasting its \textit{WRITE}. This might throttle the leader, but only if $i$ is used in the quorum to reach $Q_v$ voting weights, meaning the message of $i$ is used as the \textit{quorum formation speed determining} vote. We consider this in our calculations for a sequence of rounds by computing \textit{offsets} (additive times other replicas need to finish their consensus relative to the leader).

\begin{figure*}[t]
	\centering
	\includegraphics[width=1\textwidth]{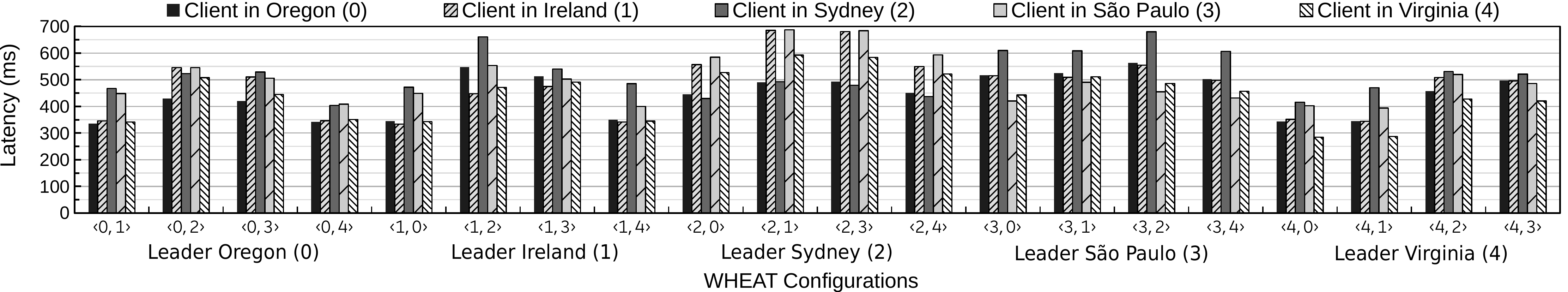}
	\caption{ Measured average request latency of 11th to 90th percentile across clients in different regions.}
	\label{fig:clientsLatencies}
\end{figure*}
\begin{figure*}[t]
	\centering
	\includegraphics[width=\textwidth]{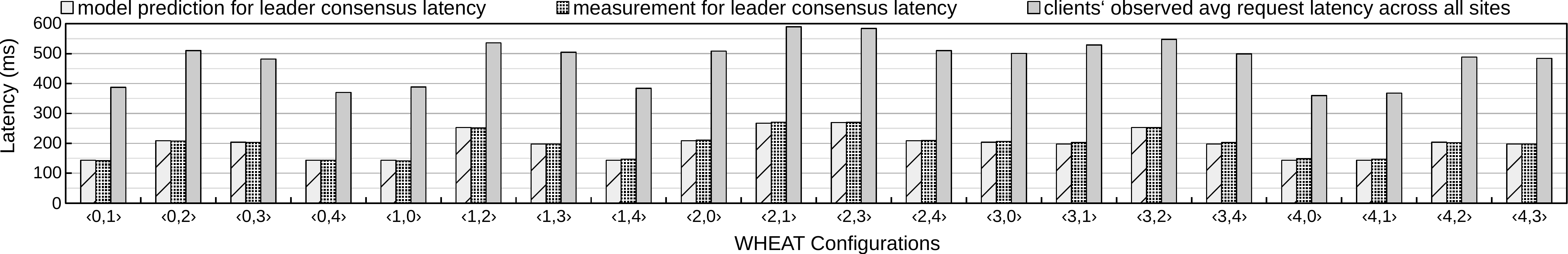}
	\caption{Comparison between predicted consensus latency,  measured consensus latency and clients' request latency.}
	\label{fig:predictionsVSmeasurements}
\end{figure*}

An AWARE configuration defines the weight configuration and selects a leader. Hence,
the number of possible configurations is the number of weight configurations multiplied with the number of possible leaders ($V_{max}$ replicas):
\begin{equation}
 \binom{3f+1+\Delta}{2f} \cdot  2f  =  \frac{\prod_{i=2f}^{3f+1+\Delta}i}{(f+1+\Delta)! }
 \end{equation}
 This yields $20$ possibilities for a $n=5,f=1,\Delta=1$ system and $504$ possibilities for a $n=9,f=2,\Delta=2$ system.
Traversing the entire search space of possible configurations becomes unfeasible for large $f$. However, (1)~BFT systems typically run with tens of nodes and (2)~if larger systems are needed, we can employ heuristics for approximating the optimum (e.g., \textit{simulated annealing}) using \textit{PredictLatency} to determine the goodness of a solution. Since we want AWARE to be practical in systems of larger sizes, we explain, implement, and validate such an approach in Section~\ref{larger_systems}.

\section{Evaluation}
\label{evaluation}

Throughout this section, we
(1)~experimentally quantify the margin of latency variations among different WHEAT configurations, (2)~compare our model prediction for consensus latency with real-world measurements in terms of accuracy, (3)~determine the correlation between consensus latency and measured request latency observed by clients across multiple regions, (4)~evaluate the run-time behavior of AWARE when carrying out self-optimizations, (5)~evaluate the maximum throughput of AWARE and investigate the monitoring overhead induced by \textit{DUMMY-PROPOSE}, and (6) reason about the system behavior in the presence of faulty replicas.

\textbf{Setup.} Unless stated otherwise, we use the Amazon AWS cloud to place our EC2 instances in specific regions. Since we do not have high hardware requirements for our latency experiments, we use the \textit{t2.micro} instance type, which is equipped with 1 vCPU, 1 GB of RAM and 8 GB standard SSD volume (gp2). We use WHEAT in the Byzantine fault model with $f=1$ and $\Delta=1$ additional spare replicas.
 Further, we select the (numbered) regions \textit{(0) Oregon}, \textit{(1)~Ireland}, \textit{(2)~Sydney,} \textit{(3) São Paulo} and \textit{(4) Virginia}.
In each region, we start one virtual machine (VM) to construct our world-spanning replicated system. Every VM carries a replica and a client which conduct latency measurements. \emph{Consensus latency} defines the time between a leader sending a proposal and the proposal being decided.
\emph{Request latency} is the time between a client sending a request and receiving enough replicas' responses to accept the result.
Replicas measure the average consensus latency of a $1000$ consensus instances sample. 
Clients simultaneously send at least $1000$ requests each 
and continue sending requests until each client has finished its measurements. A client request arriving at the leader replica may wait for some time until it gets included in a batch when there is currently a consensus instance running.
We use synchronous clients that wait for the result and send the next request after waiting for a random time interval between $0$ and $150~ms$. 
Further, clients compute the average latency from the 11th to 90th percentile (to mitigate the influence of outliers) of perceived request latencies.

\begin{figure*}[tb]
	\centering
	\includegraphics[width=1\textwidth]{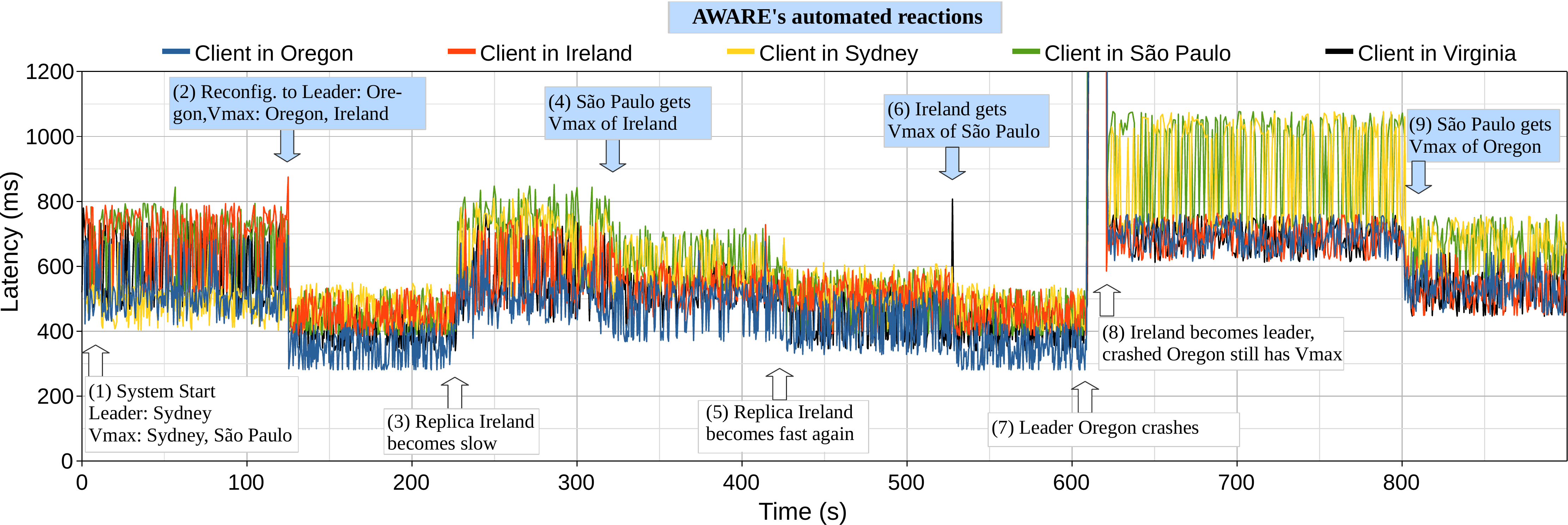}
	\caption{Runtime behavior of AWARE.}
	\label{fig:runtime}
\end{figure*}

\subsection{Margin of Latency Variations of Configurations}
\label{latency_gains}
We start by justifying the question whether a dynamic approach to self-reliantly finding a well-performing configuration is needed by showing the gap between different WHEAT configurations. Figure~\ref{fig:clientsLatencies} illustrates the observed client latencies for different regions. Each configuration is represented by a tuple $\langle L, R_{Vmax} \rangle$ where $L$ is the leader and $R_{Vmax}$ is the other replica (besides the leader) that has $V_{max}=2$ voting weight.
Each number corresponds to a region as explained in the setup and Figure~\ref{fig:clientsLatencies}.
We notice a big difference between the configurations. The best configuration is $\langle 4,	0\rangle $ showing a latency (avg. across all clients) of  $360~ms$, the left median configuration $\langle 3,4\rangle $ performs in $499~ms$ and the worst configuration  $\langle 2,1 \rangle $  requires $590~ms$. 
The best WHEAT configuration is $38.7\%$ faster than the median and $63.9\%$ faster than the worst  configuration.
Further, we make four important observations: 

 (1) \textit{Tuning voting weights can reduce latency:}  the adjustment of weights is a promising optimization to reduce the latency even if the leader is fixed (see different configurations with the same leader). 
 
(2) \textit{Leader selection may be necessary for optimal latency:} a leader in \textit{Sydney} or \textit{São Paulo} is not well connected enough to the rest of the system. Relocation can improve the latency observed by all clients.

(3) \textit{Co-located clients achieve slightly better latency:} a client co-located with the leader tends to observe lower request latencies than other clients within a specific configuration. Still, co-locating the leader with the
client does not necessarily imply a Pareto-optimum: a client in Sydney observes $492~ms$ in $\langle 2,1\rangle $ (client co-located with the leader), while it achieves its best results (among all configurations) in $\langle  0,4 \rangle $ with $403~ms$.

(4) \textit{A global optimum does not exist} but a few Pareto-optimal configurations dominate poorer performing configurations.

\subsection{Accuracy of Consensus Latency Prediction}
Our approach aims at finding a configuration with minimal leader consensus latency. Our prediction model (Algorithm~\ref{algo:PredictLatency}) lets us compute these latencies for all configurations.  

We compare our model prediction with the actual consensus latency of the leader that we measured for every configuration during our experiment (see Figure \ref{fig:predictionsVSmeasurements}). For these configurations, our predictions are off by $1.08\%$ on average. 
The highest prediction error is for $\langle 4, 0\rangle~(3.22\%)$. Since our prediction relies on latency measurements that are subject to smaller variations, we argue that these results are \textit{reasonable for choosing a well-performing configuration} -- however, AWARE might not always choose the actual best configuration but decide for some configuration that is close to the optimum. 

In our example, AWARE will pick any configuration among $\langle 0,1\rangle$, $\langle 0,4 \rangle$, $\langle 1,0\rangle$,
$\langle 1,4 \rangle$, $\langle 4,0 \rangle$, or $\langle 4, 1 \rangle$, for which it predicts a leader consensus latency of $143.5~ms$ amortized over $1000$ consensus instances. In our experiment, the measured latencies for these \textit{optimal candidates} are between $141~ms~(\langle 1,0 \rangle)$ and $148~ms~(\langle 4,0 \rangle)$. If there is an optimal configuration containing the current leader, AWARE preferably chooses it over configurations where a leader change is necessary.
On a side note, the median predicted leader's consensus latency is $202~ms~(\langle 3,4\rangle)$ and the worst is $271~ms~(\langle 2,1\rangle)$.

\subsection{Clients' Observed Request Latency}
Figure~\ref{fig:predictionsVSmeasurements} also shows the clients' observed request latency (average across all sites) for all configurations and compares them with both model predictions and measurements for consensus latency.
 As expected, consensus speed contributes to total latency. We notice a positive correlation $$\rho(L^{MP}, L^{CR}) =0.961$$  between our series (over all configurations) of model predictions for leader consensus latency $L^{MP}$ and the measurement series of average clients' request latency $L^{CR}$, indicating that \textit{faster consensus is beneficial for geographically distributed clients}.

\subsection{Runtime Behavior of AWARE}

We deploy AWARE in our usual setting and observe its behavior during the system's lifespan. Overall, the clients' request latencies show high variations, which is caused by a waiting time of a request at the leader: since  all clients simultaneously send requests and the leader batches these,  a client request may 
wait until the current consensus finishes to get into the next batch, which takes a varying time depending on how shortly the request arrived before the next consensus can be started. 
We induce events to evaluate AWARE's reactions (see Figure~\ref{fig:runtime}) to certain conditions:
\begin{enumerate}[leftmargin=16pt, label=\protect\circled{\arabic*}]
\item[\circled{1}]  Action: We start AWARE in a low-performance configuration $\langle 2,3\rangle$  with \textit{Sydney} being the leader and \textit{Sydney} and \textit{São Paulo} having maximum voting weights.
\item[\circledBlue{2}]  Reaction: After a calculation interval of $c=500$ consensus instances, AWARE decides that \textit{Oregon} and \textit{Ireland} are faster and changes its configuration to $\langle 0,1\rangle$, leading to \textit{latency gains observed by all clients across all sites}.
\item[\circled{3}]  Action: We create network perturbations, in particular we add an outgoing delay of $120~ms$ and $20~ms$ jitter to the \textit{Ireland} replica, thus making it slower 
(the client and replica of Ireland are not co-located on the same VM).
\item[\circledBlue{4}] Reaction: AWARE attributes one of the $V_{max}$ to \textit{São Paulo} while \textit{Ireland's} weight is reduced to $V_{min}$. Clients observe a small improvement in request latencies.
\item[\circled{5}]  Action: We end the network delay for \textit{Ireland}, thus the network stabilizes and the communication links of  \textit{Ireland} become just as fast as at the beginning of our experiment.
\item[\circledBlue{6}]   Reaction: AWARE notices this improvement and assigns the $V_{max}$ of \textit{São Paulo} back to \textit{Ireland}, since it predicts latency gains for this configuration. After the reconfiguration, clients observe faster request latencies identical to what happened after the first reconfiguration (Event~\circledBlue{2}).
\item[\circled{7}]  Action: We crash the leader \textit{Oregon} (which has $V_{max}$).
\item[\circled{8}]  Reaction: Replicas'  request timers expire and BFT-SMaRt triggers the leader change protocol: \textit{Ireland} becomes the next leader. Since $fV_{max}$ voting weights become unavailable, all \textit{remaining correct replicas are forced to use the same quorum} $Q_v$ (all $3$ $V_{min}$ replicas and the leader). Accordingly, clients observe higher request latencies. 
\item[\circledBlue{9}]  Reaction: AWARE redistributes the $V_{max}$ to a former  $V_{min}$ replica, \textit{São Paulo}, hence \textit{restoring some degree of variability in quorum formation}. Replicas now can form smaller quorums. This  
leads to clients observing latency improvements across all regions.
\end{enumerate}

\subsection{Maximum Throughput}
\label{throughput}
For measuring \textit{maximum throughput}, we change the instance types to \textit{c5.xlarge} (4 vCPU, 8 GB of RAM and 8 GB SSD) and use 5 VMs in our usual regions to place replicas, and 5 other VMs to launch as many clients as necessary to saturate the system. Asynchronous clients send requests of size $1$~kB after randomly waiting between $0~ms$ and $10~ms$ and replicas' responses also have a size of $1$~kB.
We compare 3 different variants: (1)~WHEAT  in a bad configuration ($\langle 2,3\rangle$), (2)~AWARE, after having adjusted to $\langle 0,4 \rangle$ using \textit{WRITE-RESPONSE}, and (3)~AWARE, in  configuration $\langle 0,4 \rangle$ and with enabled \textit{DUMMY-PROPOSE} (we bound monitoring overhead with $\omega=0.5$, hence, only every second consensus instance a replica, chosen by the rotation scheme, broadcasts a \textit{DUMMY-PROPOSE} if it's not the leader).

\begin{figure}[tb]
	\centering
	\includegraphics[width=1\columnwidth]{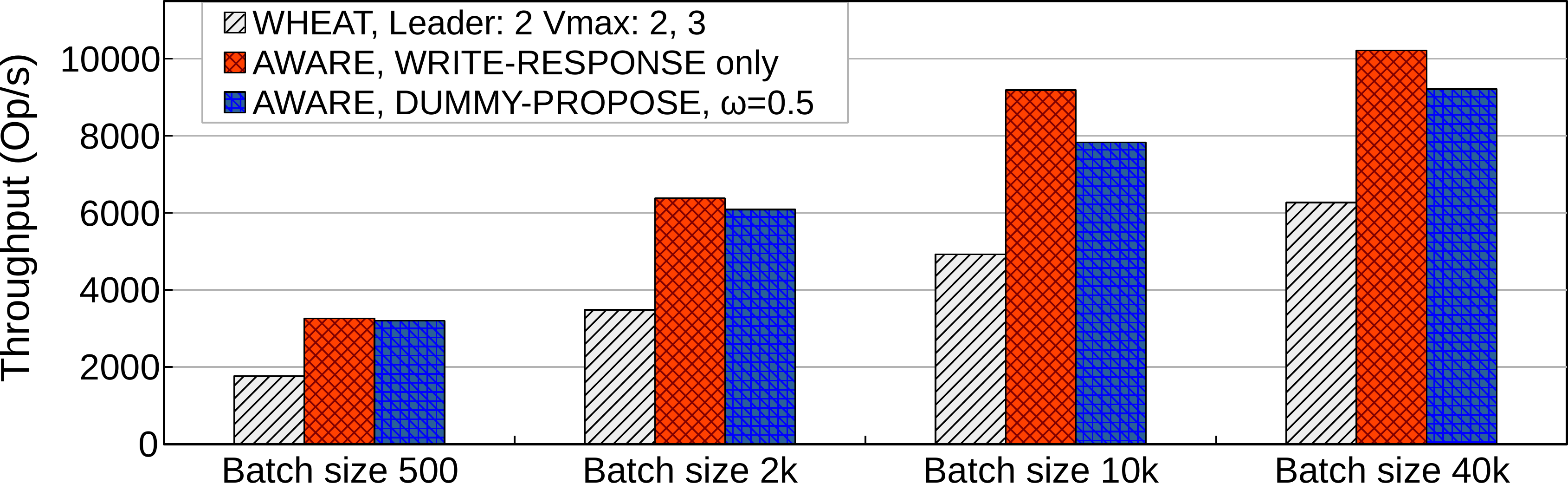}
	\caption{Maximum throughput comparison.}
	\label{fig:throughput}
\end{figure}

In principle, faster-completing consensus instances and increasing the batch size
(number of requests decided per consensus) can both raise the throughput. However, larger batch sizes also lead to higher \textit{PROPOSE} latencies and thus can slow down consensus. In our experiments, we observe (see Figure~\ref{fig:throughput}) that (1)~low consensus latency indeed has positive effects on throughput for different batch sizes and (2)~the monitoring overhead induced by enabling the \textit{DUMMY-PROPOSE} is noticeable, but still passable, given that AWARE is mainly thought of as a latency optimization technique. 



\subsection{Effect of Faulty Replicas}
\label{analysis}
AWARE deals with both crash and malicious faults as long as they are limited  to $f$ faulty replicas. In WHEAT, if replicas become unavailable for quorum formation, then their voting weights cannot be accessed until they get repaired. In AWARE, we need to distinguish two cases: first, replicas may crash, so their weight cannot be accessed for quorum formation anymore. Second, malicious replicas may adapt their behavior to prevent AWARE from redistributing weights.

\emph{Crash faults.} In the first case, AWARE detects the unavailability, as crashed replicas do not disseminate measurements nor do they respond to protocol messages. 
Thus, $+\infty$ latencies are fed into AWARE's prediction model. In case crashed replicas have $V_{max}$ voting weights, AWARE redistributes them to faster replicas (see Figure ~\ref{fig:runtime}, Event~\circledBlue{9}) and hence  restores some variability in quorum formation. If the former fast but crashed replicas are repaired and reintegrated in the system, then they might re-obtain high voting weights as well.

\emph{Malicious faults.} In the second case, malicious replicas might not participate in the consensus (e.g., do not send \textit{WRITE} messages), but still send response messages to other replicas and disseminate latency vectors that attribute them very low latencies, e.g., if $f>1$, then pairs of malicious replicas could assert that their pairwise communication links have a latency close to $0$.
Since malicious replicas might participate in the monitoring process, their unavailability cannot be easily detected by AWARE and thus restricts AWARE's ability to redistribute voting weights only between correct replicas. If we assume the worst case, then $f$ malicious replicas might in total possess up to $fV_{max}$ voting weights and force correct replicas to use the only remaining fallback quorum of $fV_{max} + (f+1+\Delta)V_{min}$ voting weights cast by all $2f+1+\Delta$ correct replicas to make progress. Still, we can always guarantee the availability of the system.
Restricting the quorum formation variability can  -- in the presence of faults -- generally happen in consensus protocols that make use of quorum systems, regardless if they use egalitarian or weighted quorums.

\subsection{Summary of Observations}

We conclude that AWARE's approach refines the practical utilization of WHEAT  in several ways: 

\emph{Ease of Deployment.} For deployment, it is irrelevant to choose a good starting configuration because AWARE provides the needed automation for finding an optimal configuration by tuning voting weights and relocating the leader.

\emph{Adjusting to Varying Network Conditions.} The quality of communication links may vary for different reasons, e.g., bad routing, overloads or DDoS attacks. This might be less a problem for Amazon's data centers but can occur for servers located in poorer connected regions. AWARE dynamically adjusts to new conditions by shifting high voting weights to replicas that are the fastest in a recent timeframe.

\emph{Compensating for Faults.}  In the worst case, $f$ replicas become unavailable. If they all have $V_{max}$ voting weights, then all correct replicas need to access the same quorum without any variability. For non-malicious behavior, AWARE detects this and restores the availability of up to $f(V_{max}-V_{min})$ voting weights in the system by redistributing high voting weights to the fastest of the remaining correct replicas. 






\section{AWARE in Larger Systems}
\label{larger_systems}
 \begin{figure*}[t]
	\centering
	\begin{subfigure}[b]{.33\textwidth}
		
		\includegraphics[width=1\textwidth]{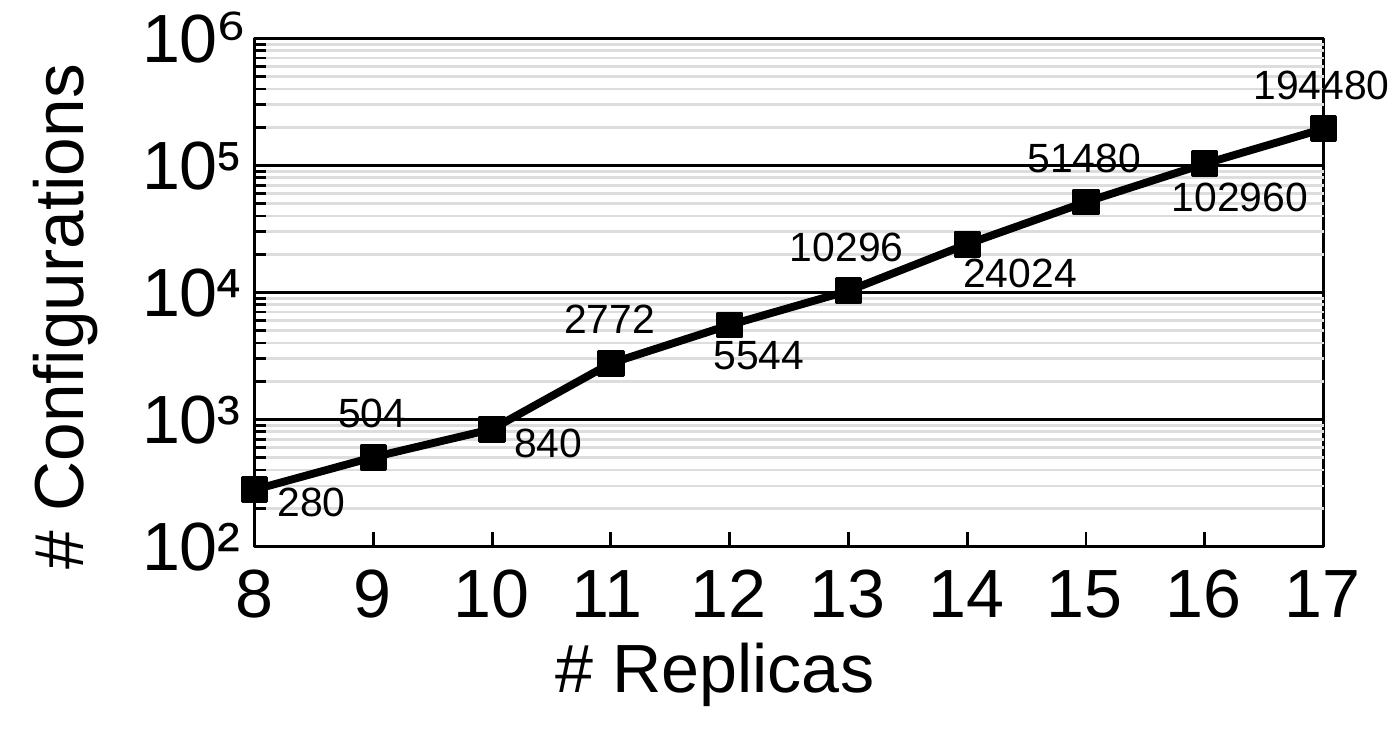}
		\caption{\# Configurations.}
		\label{fig:number-of-configs}
	\end{subfigure}
	\begin{subfigure}[b]{.33\textwidth}
	
	\includegraphics[width=1\textwidth]{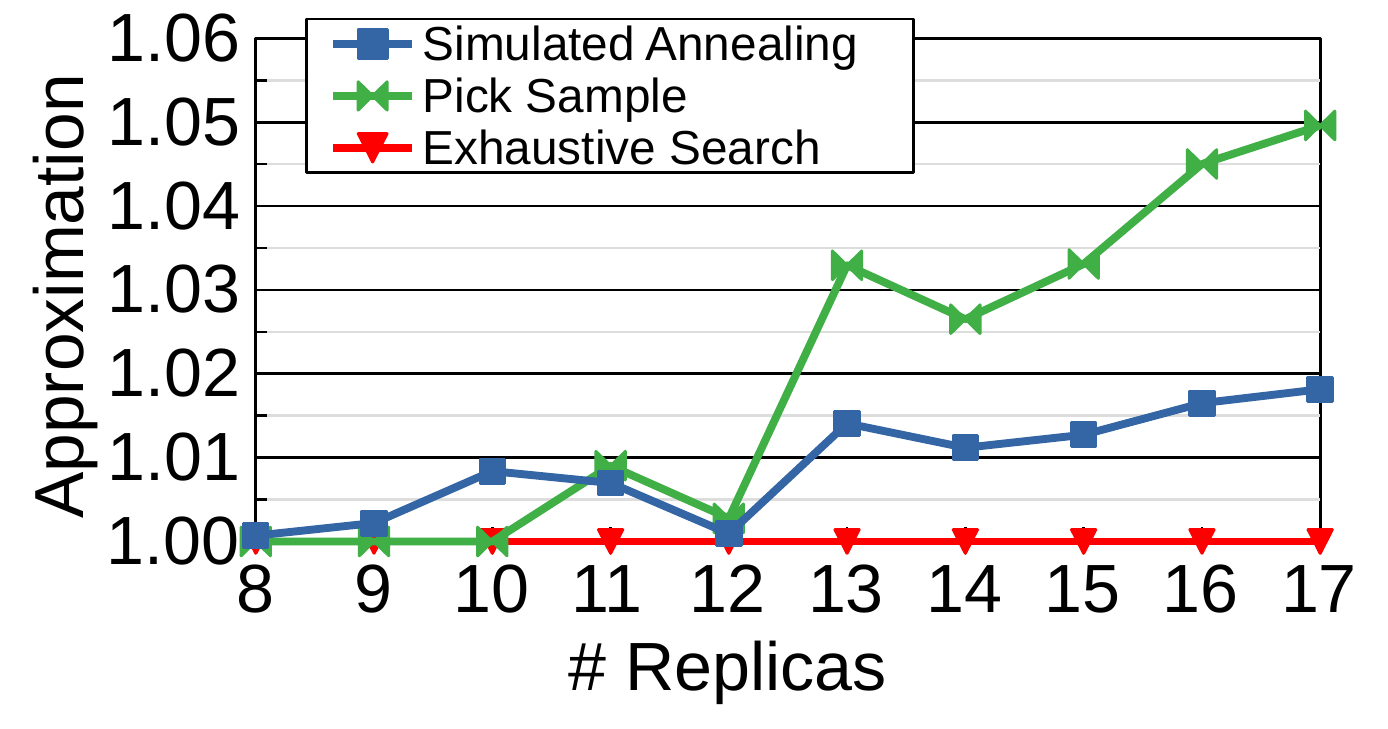}
	\caption{Approximation quality.}
	\label{prediction-accuracy-of-heuristics} 
\end{subfigure}
	\begin{subfigure}[b]{.33\textwidth}
		
		\includegraphics[width=1\textwidth]{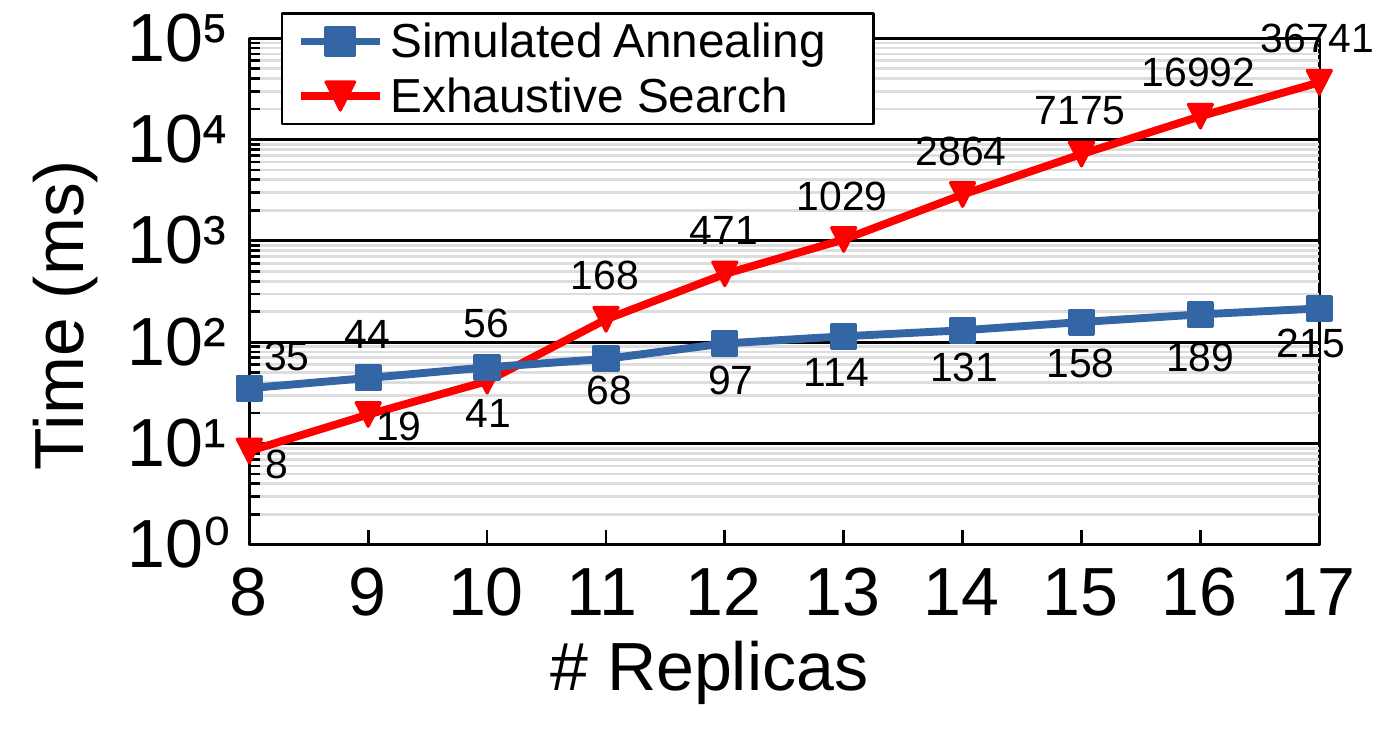}
		\caption{Computation time.}
		\label{fig:computation-time-of-configs} 
	\end{subfigure}

	\caption[]{A heuristic can help to efficiently traverse the configuration space to find a good solution.
	}
	\label{fig:prediction-larger-n} 
\end{figure*}

In this section, we want to (1)~explain how AWARE can be employed in larger system environments, which includes dealing with an exponentially growing configuration space, and (2)~conduct experiments with several, widely-spread replicas to show that adaptive weighted replication is a suitable approach for geographically-scalable BFT consensus.
In Section~\ref{latency-prediction} we stated that the fast-growing configuration space of possible weight distributions quickly makes exhaustive search impossible. Exhaustive search predicts the latency for all possible configurations. The problem is that the number of weight configuration grows fast because these are essentially combinations for selecting $2f$ replicas with weight $V_{max}$ out of a total of $3f+1+\Delta$ replicas in the system. Figure~\ref{fig:number-of-configs} shows the growing number of configurations for exemplary system sizes depicted in Table~\ref{table:example-system-sizes}.

 \begin{algorithm}[t] 
 	\footnotesize
 	\caption{\emph{SimulatedAnnealing} is a heuristic for efficiently traversing the search space of configs $C$}
 	\DontPrintSemicolon
 	\KwData{replica set $I$, system sizes $n$, $f$, $u$, $\Delta$, latency matrices for \textit{PROPOSE} $\hat{M}^{P}$ and \textit{WRITE}  $\hat{M}^{W}$,  consensus id $cid$, start temperature $t_0$, cooling rate $\theta$,  temperature $threshold$}
 	\KwResult{best (approx.) performing configuration found}
 	\label{algo:SimulatedAnnealing}
 		\tcc*[h]{Initially, replicas deterministically start the search with the same, current config.}  \;
 	$c \leftarrow $ current config $c_0 \in C$ \;
 	$c.prediction \leftarrow predictLatency(I, c, \hat{M}^{P}, \hat{M}^{W}, n, f, \Delta); $  \;
 	$c_{approx} \leftarrow c$  \;
 	$temp \leftarrow t_0 $ \;
 	$random \leftarrow$ new Random(cid) \;
 	
 	\While{$temp > threshold$} {    
 		\tcc*[h]{Assign a $V_{max}$ to another replica}  \;
 		$replicaFrom \leftarrow c.R_{max}[random.nextInt(u)] $ \;
 		$replicaTo \leftarrow c.R_{min}[random.nextInt(n-u)] $ \;
 		$c' \leftarrow c.swap(replicaFrom, replicaTo) $ \;
 		\If{replicaFrom is leader}{
 			$c'.setLeader(replicaTo)$ \;
 		}
 		$c'.prediction \leftarrow predictLatency(I, c', \hat{M}^{P}, \hat{M}^{W}, n, f, \Delta); $  \;
 		\tcc*[h]{If new solution is better, accept it}  \;
 		\eIf{c'.prediction < c.prediction}{
 			$c \leftarrow c' $ \;
 		}{
 			\tcc*[h]{Compute an acceptance probability}  \;
 			$rand \leftarrow random.nextDouble()$ \;
 			\If{$exp(\frac{-(c'.prediction - c.prediction)}{temp})> rand$}{
 				$c \leftarrow c' $ \;
 			}
 		}
 		\If{$c'.prediction < c_{approx}.prediction$}{
 			$c_{approx} \leftarrow c'$ \;
 		}	
 		\tcc*[h]{Cool down the system}  \;
 		
 		$temp \leftarrow temp \cdot (1 - \theta) $ \;
 	}
 	\textbf{return} $c_{approx}$ \;
 \end{algorithm}
 
 \begin{table}[]
 	\centering
 	\resizebox{\columnwidth}{!}{%
 		\begin{tabular}{|c|c|c|c|c|c|c|c|c|c|c|}
 			\hline
 			$n$    & 8 & 9 & 10 & 11 & 12 & 13 & 14 & 15 & 16 & 17 \\ \hline
 			$f$     & 2          & 2          & 2           & 3           & 3           & 3           & 4           & 4           & 4           & 4           \\ \hline
 			$\Delta$ & 1          & 2          & 3           & 1           & 2           & 3           & 1           & 2           & 3           & 4           \\ \hline
 		\end{tabular}%
 	}
 	\caption{Exemplary system sizes.}
 	\label{table:example-system-sizes}
 \end{table}

\subsection{Simulated Annealing} To address this problem, we make use of a heuristic to efficiently traverse the search space. \textit{Simulated annealing}~\cite{Kirkpatrick671} is a technique for finding an approximation of the global optimum of some function (see Algorithm~\ref{algo:SimulatedAnnealing}). It is essentially a local search that aims for improving on a current solution~$c$ by randomly modifying it slightly (which is called neighbor picking), yielding a new configuration $c'$. If the new configuration $c'$ improves the function output (in our case has a lower predicted latency), then the simulated annealing algorithm uses $c'$ to proceed with its search. However, if $c'$ worsens the function output, it can still be used to proceed with some acceptance probability which depends on the current \textit{temperature} (a monotonously decreasing value) and the energy of the solution, which is essentially the difference between the predicted latency of $c'$ and $c$.  By accepting a temporary worsening, simulated annealing can \textit{jump}, that is, escape from a local optimum to find the global optimum. However, these jumps become less frequently with decreasing temperature. The exit condition for this kind of search can be implemented, e.g., by using a temperature threshold. This guarantees termination if the temperature decreases at a fixed rate. Since simulated annealing is a probabilistic algorithm, it needs to generate a sequence of random numbers during its execution. Because we want to guarantee that replicas deterministically find a consistent solution, we need to employ a pseudo random number generator \textit{PRNG(s)} that generates the same sequence of random numbers in all replicas given the same input seed~$s$. We let replicas use the consensus id (replicas decide on a possible reconfiguration at specific consensus instances) as seed for generating these numbers. For simulated annealing, we use the following parameters:  $t_0=120$, $\theta=0.0055$, $threshold=0.2$ (which were empirically chosen). With these parameters, the algorithm terminates after probing $1160$ configurations.

\begin{figure*}[t]
	\centering
	\begin{subfigure}[b]{0.33\textwidth}
		
		\centering
		\resizebox{\textwidth}{!}{%
			\begin{tabular}{|c|c|c|}
				\hline
				& initial replica set                                                                                                                        & $\Delta$-replicas                                                                                       \\ \hline
				$f=2$ & \begin{tabular}[c]{@{}c@{}}Mumbai, São Paulo,\\  Paris, Ohio, Sydney, \\ Tokyo, Canada\end{tabular}                                & \begin{tabular}[c]{@{}c@{}}Virginia, \\ Ireland, \\ Stockholm, \\ Frankfurt\end{tabular} \\ \hline
				$f=3$ & \begin{tabular}[c]{@{}c@{}}Mumbai, São Paulo, \\ Paris, Ohio, Sydney, \\ Tokyo, Canada, Bahrain,  \\ Stockholm, Seoul\end{tabular} & \begin{tabular}[c]{@{}c@{}}Frankfurt, \\ Virginia, \\ Ireland, \\ Oregon\end{tabular}    \\ \hline
			\end{tabular}%
		}
		
		\caption{Test environments.}
		\label{table:replica-regions}
	\end{subfigure}
	\begin{subfigure}[b]{0.33\textwidth}
		
		\includegraphics[width=\textwidth]{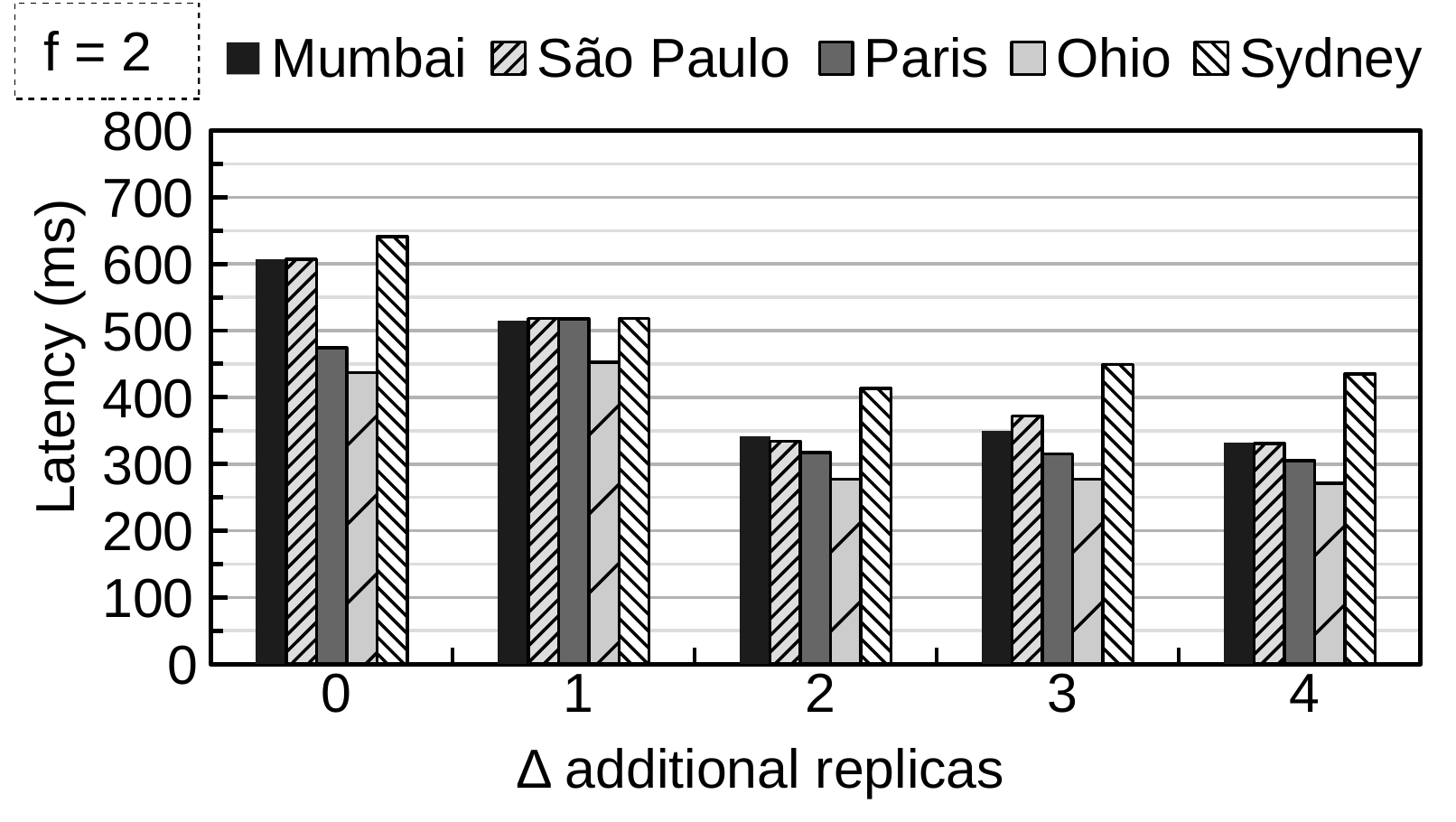}
		\caption{$f=2 $ configuration.}
		\label{fig:f2_measurements}
	\end{subfigure}
	\begin{subfigure}[b]{0.33\textwidth}
		
		\includegraphics[width=1\textwidth]{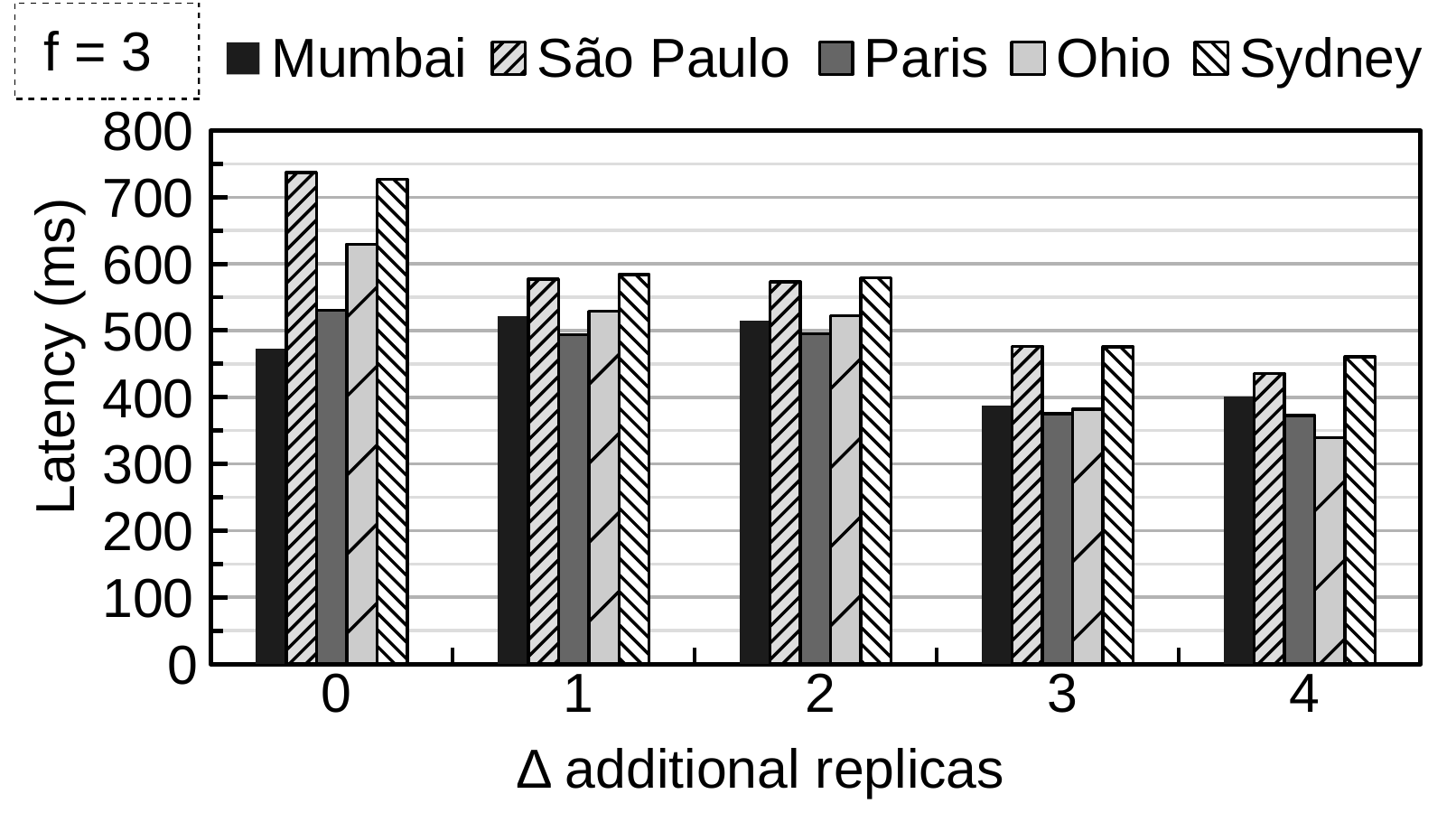}
		\caption{$f=3$ configuration.}
		\label{fig:f3_measurements} 
	\end{subfigure}
	
	\caption[]{Latency results of AWARE experimentally tested in larger-scale setups.
	}
	\label{fig:larger-n} 
\end{figure*}

\subsection{Approximation Quality}
We validate the quality of our heuristic \textit{Simulated Annealing} (SA) (using above parameters) by comparing the predicted latency of the approximate solution it finds to the optimum found by \textit{Exhaustive Search} (ES) and a third strategy, \textit{PickSample}, which samples through the configuration space in equal steps and probes the same number of configurations as SA does. We compute the average deviation from the optimum by simulating a set of $1000$ randomly generated setups for each system size $n$. 
Figure~\ref{prediction-accuracy-of-heuristics} shows that the predicted latencies of solutions found by SA are on average less than $1.02$ higher than the optimum found by ES. It can be observed that SA performs better than naive sampling as solutions found by \textit{PickSample} display a higher derivation from the optimum (up to $1.05$).

\subsection{Computation Time} Further, we evaluate the average time a replica needs to find a solution on an 
Intel i7-4790 @3.60 GHz processor using $1000$ randomly generated setups for simulation for each system size $n$.
Figure \ref{fig:computation-time-of-configs} compares the time needed to find a solution between two strategies, SA (with the above specified parameters) and ES for different system sizes (see Table~\ref{table:example-system-sizes}). As expected the time needed by ES grows exponentially for an increasing $n$. SA only computes through a constant number of configurations, however, the time still increases as the \textit{PredictLatency} function needs more time to predict the latency of larger systems, since it essentially is a simulation of the consensus protocol run. To be precise, the time complexity of \textit{PredictLatency } is $\mathcal{O}(n^2log(n))$ because every replica maintains a priority queue of messages (ordered by ascending arrival time) as a min heap, which takes $\mathcal{O}(nlog(n))$ time to construct, and we need to consider each of the $n$ replicas during the simulation. The constant number of configurations that SA examines depends on the chosen parameters for start temperature, temperature threshold, and the cooling schedule.
In total, the time complexity of SA is in $\mathcal{O}(n^2log(n))$.

%
%

\subsection{Experiments on AWS}

In this section,  we describe experiments on the Amazon AWS cloud infrastructure to evaluate how adaptive weighted replication behaves for systems with larger replica sizes than the $f=1, n=5$ setup, which we evaluated in Section~\ref{evaluation}. In particular, we want to investigate if for a fixed $f$, adding additional spare replicas (called $\Delta$-replicas) to an initial replica set can lead to latency gains observed by clients spread across the world.

\subsubsection{Setup} We use the Amazon AWS infrastructure and the same overall setup as in Section~\ref{evaluation} to do our measurements but add more regions to place the replicas (see Table~\ref{table:replica-regions}). Here, for both $f=2$ and $f=3$ we start with an initial replica set and then repeat the experiment with increased system sizes where the added spare replicas serve as $\Delta$-replicas used to improve the consensus latency of the system. These replicas are added to the system configuration in the order in which they appear in Table~\ref{table:replica-regions}. Replicas use AWARE  to self-optimize. Even for a $\Delta=0$ configuration, where all replicas have equal weight, AWARE still automates the selection of the leader, hence choosing the best leader for the given environment. AWARE is configured to self-optimize after every $500$ consensus instances. 
Moreover, we locate 5 clients in different regions across the globe, namely in \textit{Mumbai},  \textit{São Paulo}, \textit{Paris}, \textit{Ohio}, and \textit{Sydney}. Each system environment is tested by
clients first sending $500$ requests for a warm-up phase to make sure AWARE has adapted to a fast-performing configuration.
Then, clients measure request latency as the median of $1000$ observed requests which they simultaneously send.

\subsubsection{Observations} Figure~\ref{fig:f2_measurements} and Figure~\ref{fig:f3_measurements} show the results for the $f=2$ experiments with system sizes between 7 and 11 replicas and the $f=3$ experiments with system sizes between 10 and 14 replicas, respectively. For the $f=2$ experiments we observe that after adding \textit{Virgina} as additional replica, the average request latency over all clients improves from $553~ms$ to $505~ms$, and further adding \textit{Ireland} to the system results in an average of $337~ms$. Adding even more replicas ($\Delta=3$ or $\Delta=4$) does not yield further substantial gains. Moreover,  for the $f=3$ experiments we observe that after adding \textit{Frankfurt}, the average latency of all clients improves from $619~ms$ to $541~ms$. Interestingly, the latency that the client in Mumbai observes slightly increases. This is due to the fact that Mumbai is leader in the $\Delta=0$ configuration, but another replica becomes leader in the $\Delta=1$ configuration and co-residency with the leader is beneficial for a client. As we add more replicas, the average latency across clients improves further, in particular $537~ms$ (adding \textit{Virginia}), $419~ms$ (adding \textit{Ireland}) and $402~ms$ (adding \textit{Frankfurt}). We conclude that adaptive weighted replication can improve the latency of geographically-scalable BFT consensus, as assigning high voting weights to fast replicas supports the emergence of fast quorums in the system. Fast consensus generally is beneficial for clients, however other factors also exist, e.g., being located near the protocol leader.


\section{Integration of AWARE into the Hyperledger Fabric Blockchain Platform}
\label{hyperledger}

To illustrate a practical use case for the AWARE protocol, we show that it can be employed as a building block for distributed ledger infrastructures. As an example, we integrate AWARE into the Hyperledger Fabric (HLF)~\cite{androulaki2018hyperledger} blockchain platform as a self-adapting ordering service. Specifically, its task is to establish agreement on which block is appended next to the blockchain. Our protocol is particularly tailored as consensus substrate for blockchain infrastructures that (1)~are geographically decentralized with ordering nodes being spread across different regions in the wide-area network, (2)~adopt Byzantine fault-tolerance, and (3)~want to achieve adaptiveness to their environments.

\subsection{Hyplerledger Fabric} 
Hyperledger Fabric~\cite{androulaki2018hyperledger} is a modular and extensible open-source 
blockchain platform that assumes the permissioned blockchain model.  Its core innovation is to provide an abstraction and separation of different concerns, which manifest in distinct building blocks. The \textit{ledger} is a totally ordered, append-only blockchain maintained by the \textit{endorsing peers} that execute transactions, thus generating \textit{endorsements} (result of an execution against the current state, specifically the read and write
sets). Clients verify and assemble endorsements into signed \textit{envelopes}, which contain the endorsing peers' read and write sets, and submit these envelopes to a separate ordering service.
The stateless \textit{ordering service} is used to atomically broadcast blocks to endorsement peers. This service creates a total order of transactions by running consensus among the \textit{ordering nodes}, which create a sequence of blocks of ordered envelopes. Each newly created block is then disseminated to the receiving endorsing peers, which append it to the ledger. 
In HLF, the notion of membership is flexible: a \textit{membership service provider} manages the mapping of node identities to their public keys. 
Fabric introduces the \textit{execute-order-validate} paradigm, which separates the transaction flow into (1) confirming the correctness of a transaction, executing it, and outputing an endorsement, (2)  using a consensus protocol to order transactions by reaching agreement on the succession of the blocks, and (3) validating a transaction, e.g., against application specific trust assumptions.

\subsection{Integration of AWARE into Hyperledger Fabric} 
Sousa et al. designed and implemented a BFT ordering service for HLF on top of BFT-SMaRt~\cite{sousa2018byzantine}. We use the same service to integrate AWARE, which uses interfaces identical to that of BFT-SMaRt. HLF provides a client interface to submit envelopes to an external ordering service (HLF \textit{consenter}). These are submitted to a Java \textit{frontend}, which consists of a client thread pool, receiver thread, and an asynchronous \textit{BFT Proxy} (which is part of the client-side BFT-SMaRt library), which invokes these envelopes: they are broadcasted to the ordering nodes and subsequently ordered as they pass through the \textit{total order multicast layer} of BFT-SMaRt. The ordering nodes extend BFT-SMaRt's \textit{ServiceReplica} class. They receive a stream of totally ordered envelopes and use a \textit{blockcutter} to aggregate them into blocks. Created and signed blocks are distributed back to the \textit{BFT Proxy} of all receiving (listening) frontends by utilizing the \textit{Replier} interface of BFT-SMaRt. Once the BFT Proxy has gathered sufficiently many equal and verified messages for a block, the block can be passed to HLF to be appended to the ledger.

\subsection{Experimental Setup} 
In our experiments, we want to evaluate the AWARE ordering service for HLF, in particular, the latencies for ordering envelopes, generating blocks, and receiving them back as observed by frontends located in different AWS regions. We use a $f=1$ and $\Delta=1$ configuration and  choose the regions \textit{Sydney (leader, V=2)}, \textit{São Paulo (V=2)}, \textit{California (V=1)}, \textit{Tokyo (V=1)} and \textit{Stockholm (V=1)} to place a \textit{t2.micro} instance. Each instance runs both an ordering node and a frontend. An extra frontend runs in \textit{Seoul}. Frontends send sufficiently many envelopes (with transaction message payload of size 100 bytes) %
to satisfy a throughput of at least 100 envelopes/s in the system. The ordering service is configured to create blocks containing 10 envelopes and distribute created blocks to all frontends. Note that a consensus instance may contain multiple blocks. The AWARE ordering nodes use a calculation interval of $c=250$ after which they may change the configuration to optimize the consensus latency.
 
  \begin{figure}[t]
 	\centering
 	\begin{subfigure}[b]{.49\textwidth}
 		\includegraphics[width=1\columnwidth]{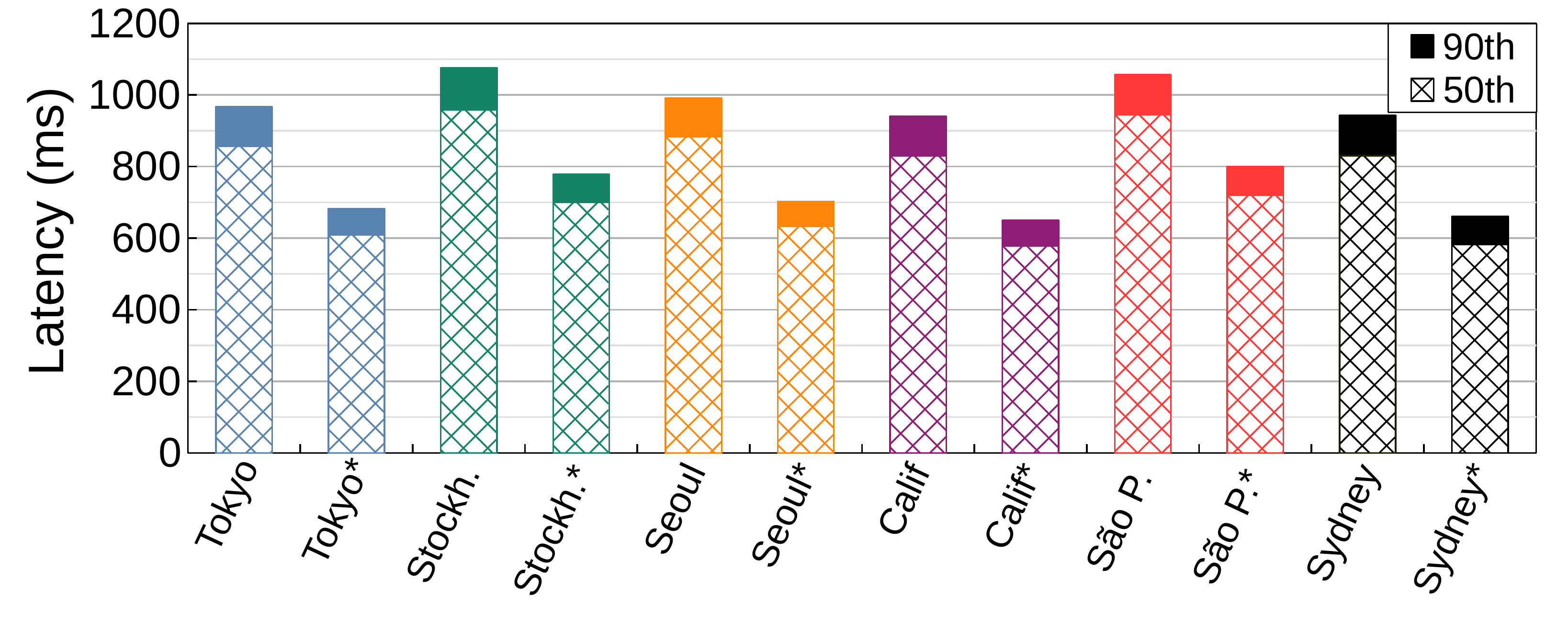}
 		\caption{Latency across frontends before and after* optimization.}
 		\label{fig:frontends:all} 
 	\end{subfigure}
 	 	\begin{subfigure}[b]{.49\textwidth}
 		\includegraphics[width=1\columnwidth]{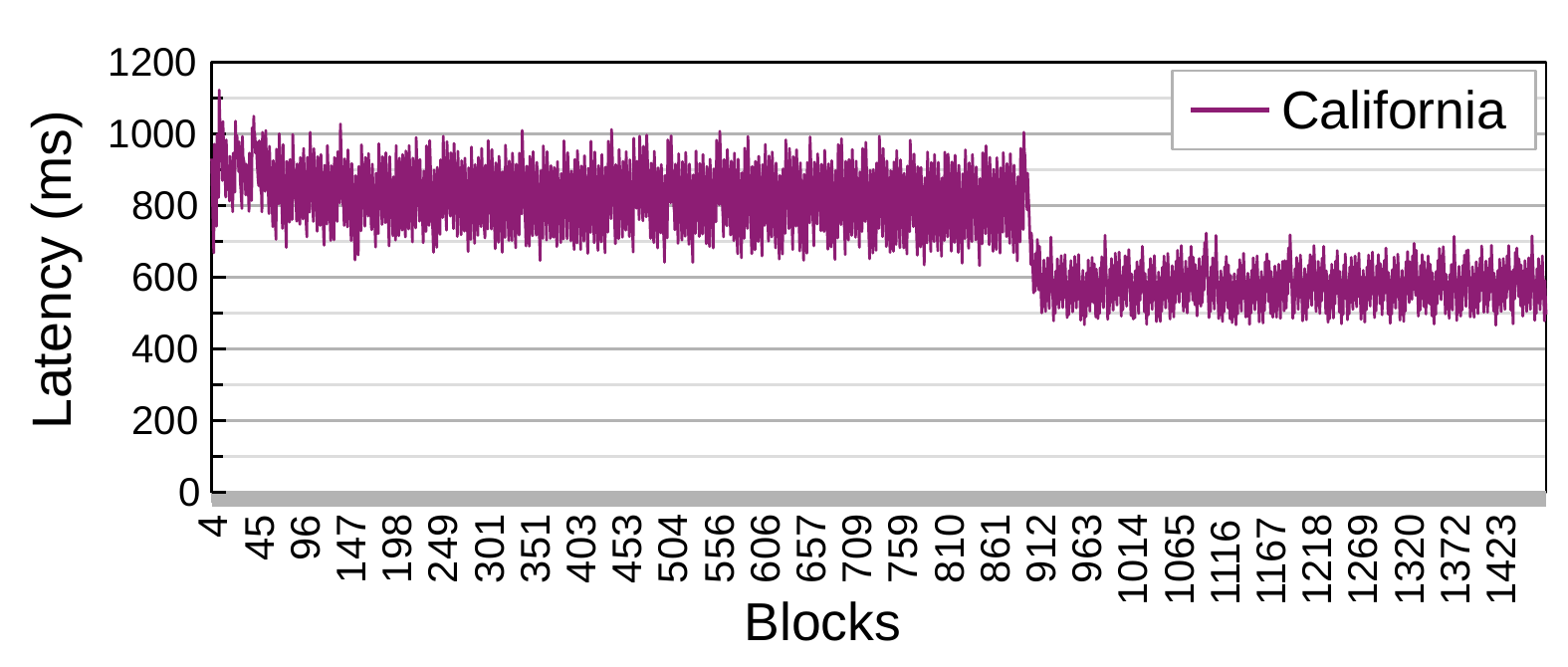}
 		\caption{Frontend in California.}
 		\label{fig:frontend:california}
 	\end{subfigure}
 	\caption{AWARE as self-adapting ordering service for Hyperledger Fabric leads to latency gains observed by frontends. }
 	\label{fig:evaluation-hyperledger} 
 \end{figure}

\subsection{Observations}
We show the envelope latencies as observed by all of the frontends in Figure~\ref{fig:frontends:all}.  Roughly at around block \#900,\footnote{Note that blocks and consensus decisions are different things. The reconfiguration around block 900 is due to the the reconfiguration triggered after 250 consensus
	instances.} AWARE reconfigures the system, shifting the maximum voting weight $V_{max}$ from \textit{São Paulo} to \textit{California}. 
 The runtime behavior from \textit{California's} perspective is shown in Figure~\ref{fig:frontend:california}.
This optimization results in a latency improvement observed by all frontends (see Figure~\ref{fig:frontends:all} for median and 90th percentile request latencies), e.g., the median latency observed by \textit{Sydney} (where the replica stays leader) decreases from $830~ms$ to $579~ms$, which corresponds to a speedup of $1.43$.
 AWARE predicts the improvement of the consensus latency from $323~ms$ to $180~ms$ after the reconfiguration. Note that latency gains that frontends observe are higher than this $143~ms$ consensus latency decrease. This is because an envelope received by the ordering service might not immediately be proposed by the leader, since the currently running consensus instance needs to be finished first. This results in a random waiting time of the envelope at the leader, whereby the expected waiting time is roughly half of the consensus latency.

\section{Related Work}
\label{related-work}
A variety of research touches the fields of  SMR optimizations in WAN environments \cite{amir10steward, Mencius, mao2009towards, EBAWA, moraru2013there,  wei2013fast} and dynamic approaches for latency awareness\cite{liu2017leader, DBLP:conf/dsn/EischerD18, Carvalho18:Dynamic,meredith09when}. Further, recent research efforts develop and refine BFT consensus protocols for blockchains~\cite{miller2016honey, kogias2016enhancing, gilad2017algorand, Kogias2018omniledger, li2018gosig,  liu2018scalable,  hotstuff19, losa2019stellar, bessani2020byzantine} or integrate these protocols into existing blockchain platforms such as Hyperledger Fabric~\cite{androulaki2018hyperledger, sousa2018byzantine, ruesch2019bloxy}.

\subsection{WAN Optimizations for SMR} \textit{Steward}~\cite{amir10steward} is a hierarchical architecture for scaling BFT replication in WANs. Multiple replication groups (each group is located at a site and runs Byzantine agreement) are geographically distributed and groups are connected over a CFT protocol. 
 Like our approach, it uses additional spare replicas but with much higher replication costs ($4$ replicas are needed at each site).
\textit{Stewards}'s idea of a hierarchical architecture is also used in the recent \textit{Fireplug}~\cite{neiheiser2020fireplug} for efficient geo-replication
of graph databases by compositing multiple BFT-SMaRt groups.
\textit{Mencius}~\cite{Mencius} is an approach for building efficient SMR for WANs by employing a rotating coordinator scheme 
where clients choose their geographically closest replica as the coordinator. However, \textit{Mencius} only supports the CFT model because of its skipping technique.
The idea of a rotating leader in \textit{Mencius} is later enhanced in the \textit{RAM} protocol~\cite{mao2009towards}, which additionally employs attested append-only
memory and assumes \textit{mutually suspicious
domains} to achieve low latency SMR for uncivil WANs. In our work, we assume the BFT model where clients do not trust their local leaders. 
\textit{EBAWA}~\cite{EBAWA} is a protocol that improves SMR in WANs under a hybrid fault model. A trusted component on the replicas allows reducing the number of replicas in the system to $2f+1$ and  the communication steps needed for agreement to~$2$. 
It also uses a rotating leader technique where clients send their requests to their local server.
In contrast, \textit{Egalitarian Paxos}~\cite{moraru2013there} allows all replicas to propose 
and employs a mechanism for solving conflicts if operations interfere. Clients choose a well-connected replica to propose their operations.  \textit{GeoPaxos}~\cite{coelho18geopaxos} decouples order from execution, utilizes partial order instead of total order and exploits geographic locality to achieve fast geographic SMR, but only tolerates crash faults just like Egalitarian Paxos.
Another recent crash-fault tolerant wide-area replication optimization is \textit{Weave}~\cite{eischer2020low}, which can give fast execution
\textit{guarantees} for client requests by avoiding that clients need to wait for
wide-area communication steps. \textit{Weave} employs local groups (of $f+1$ replicas)  that assign sequence numbers to the write requests of their corresponding local clients.
Further, it is also possible to assist system integrators with geographic SMR deployments: a recently proposed method~\cite{numakura2019evaluation}  ranks possible SMR replica
deployments  by calculating expected performance from known round trip times between replicas.

\subsection{Dynamic Approaches} The \textit{Droopy} and \textit{Dripple} protocols~\cite{liu2017leader} follow a dynamic approach to reduce latency for geo-replicated state machines under imbalanced localized workloads. The authors suggest that the choice (and the number) of leaders depends on both replica configuration and workload, thus is subject to variations over time. \textit{Droopy} dynamically reconfigures the leader set of each partition while \textit{Dripple} coordinates state partitions. Further research work shows that clients can also dynamically react to changing workloads by efficiently changing their quorum selections to achieve good performance~\cite{meredith09when}.
A protocol for latency-aware leader selection in geo-replicated systems is \textit{ARCHER}~\cite{DBLP:conf/dsn/EischerD18}, which uses clients' observed end-to-end response latencies to select the optimal leader and hence can dynamically adjust to varying workloads. In contrast, AWARE measures replica-to-replica latencies and uses weight tuning in addition to leader selection. We follow the empirical observations of WHEAT, in which a mechanism that makes clients closer to their leaders (or using a leader in the same region) gives less latency gains than just using the fastest replica as the leader~\cite{sousa2015separating}.
Dynamic adaptation of consensus algorithms for BFT systems was also studied in Carvalho et al.~\cite{Carvalho18:Dynamic} by the implementation of 
switching algorithms in the BFT-SMaRt library.

\subsection{BFT Consensus in Blockchains}
A number of BFT consensus protocols  has been developed for blockchains because a  \textit{"one size fits all solution"} may be impossible to design~\cite{singh2008bft, androulaki2018hyperledger, duan2018beat} which leads to BFT consensus protocols being tailored to specific purposes, as they differ in their ambitions and assumptions~\cite{cachin2017blockchain, bano2017consensus, berger2018scaling, ruesch2019bloxy}. For instance, \textit{Algorand}~\cite{gilad2017algorand} scales for a large number of nodes and was evaluated with up to 500k nodes while \textit{FastBFT}~\cite{liu2018scalable} employs trusted hardware components to achieve low latencies. \textit{HoneyBadgerBFT}~\cite{miller2016honey} tolerates arbitrary
network failures, as it adopts asynchronous atomic broadcast.
\textit{Gosig}~\cite{li2018gosig}  is able to
withstand adversarial network conditions on a wide-area network. 
This variety is respected by \textit{Hyperledger Fabric}'s modularity~\cite{androulaki2018hyperledger} as it encapsulates consensus in a distinct building block called the \textit{ordering service}, ideally allowing operators to choose the consensus
 that fits best for their application. However, in practice, integrating a BFT consensus
protocol still requires substantial changes (e.g., for majority voting) thus breaking Fabric’s modularity. \textit{Bloxy}~\cite{ruesch2019bloxy} tackles this problem: it is a blockchain-aware trusted
proxy that encapsulates the client functionality
of BFT consensus and thus simplifies and accelerates the integration of BFT
consensus protocols into HLF. 




\section{Conclusion}
\label{conclusion}
World-spanning Byzantine consensus systems can benefit from dynamic self-optimizing approaches. We showed how to construct such a dynamic approach using WHEAT as underlying weighting scheme and BFT-SMaRt as replication protocol by letting the system perform continuous measurements and decide on an optimal configuration. Further, we described a deterministic, self-optimization algorithm that enables the system to minimize its consensus latency and thus to faster respond to clients. 

AWARE\footnote{The open-source implementation of AWARE is available at \\ \url{https://github.com/bergerch/aware}.} enriches the idea of weighted replication by providing the needed automation to adapt to changing environmental conditions.
Our method implements resilient, \textit{adaptive} Byzantine consensus. In particular, it automates voting weights tuning and leader positioning, hence thriving for latency gains at run-time by selecting a fast performing WHEAT configuration.
Evaluation results with several AWS EC2 regions show that the potential for latency and throughput gains is substantial. 
 For many blockchains seeking a high degree of decentralization,  geographical scalability (where many nodes are widely spread across the globe) becomes an important limitation. 
We demonstrated that AWARE also works with larger system sizes and can be used as a self-adapting consensus substrate for blockchain infrastructures, such as Hyperledger Fabric.

\ifCLASSOPTIONcompsoc
  \section*{Acknowledgments}
\else
  \section*{Acknowledgment}
\fi

This work was funded by the Deutsche Forschungsgemeinschaft (DFG, German Research Foundation) – 268730775 (OptScore) and by FCT through projects IRCoC (PTDC/EEI-SCR/6970/2014), ThreatAdapt (FCT-FNR/0002/2018), and the LASIGE Research Unit (UIDB/00408/2020 and UIDP/00408/2020).

\ifCLASSOPTIONcaptionsoff
  \newpage
\fi



%



\bibliographystyle{IEEEtran}
\bibliography{IEEEabrv,bibliography}

%

\vskip -1em
	\begin{IEEEbiographynophoto}{Christian Berger} 
received the BSc and MSc degrees in computer science from the University of Passau, Germany, in 2014 and 2017, respectively. In 2018, he started working as a research associate and PhD candidate at the assistant professorship of security in information systems. His research interests include Byzantine fault tolerance, resilient distributed systems as well as blockchain and distributed ledger technology.
\end{IEEEbiographynophoto}
\vskip -1em
\begin{IEEEbiographynophoto}{Hans P. Reiser}
is an assistant professor at the University of Passau, Germany, associated with the
Institute of IT Security and Security Law (ISL).  He graduated in computer science at University of Erlangen 
and he holds a PhD (Dr. rer. nat.) in the area of middleware for fault-tolerant systems from Ulm University. He spent
one semester at the Carnegie Mellon University, Pittsburgh, USA as a visiting professor, and one semester at
Karlsruhe Institue of Technology as visiting full professor.  Hans P. Reiser’s research focus is on technical
aspects of reliability and security in distributed systems.   More information about
him can be found at \url{https://www.fim.uni-passau.de/sis/}.
\end{IEEEbiographynophoto}
\vskip -1em
\begin{IEEEbiographynophoto}{João Sousa}
	obtained his graduation at Faculdade de Ciências, Universidade de Lisboa (FCUL). In 2009 he joined LASIGE as a junior researcher and became the leader programmer for the BFT-SMaRt replication library. After finishing his masters degree in 2010, João started his PhD in 2011 (also at LASIGE/FCUL). He delivered his PhD thesis in 2017, and defended it in 2018. In 2019, he was awarded the William C. Carter PhD Dissertation Award in Dependability. Currently, he remains in LASIGE as a researcher.
\end{IEEEbiographynophoto}
\vskip -1em
\begin{IEEEbiographynophoto}{Alysson Bessani}
	is an Associate Professor of the Department of Informatics of the University of Lisboa Faculty of Sciences (Portugal), and a member of LASIGE research unit. He received his B.S. degree in Computer Science from UEM (Brazil) in 2001, the MSE and PhD in Electrical Engineering from UFSC (Brazil) in 2002 and 2006, respectively. He spent some time as a visiting professor in Carnegie Mellon University (2010) and as a visiting researcher in Microsoft Research Cambridge (2014). Alysson participated in more than ten international projects and co-authored more than 100 peer-reviewed publications on dependability, security, and distributed systems. Additional information about him can be found at \protect\url{http://www.di.fc.ul.pt/~bessani}.
\end{IEEEbiographynophoto}







\end{document}